\documentclass[preprint,superscriptaddress,showpacs]{revtex4-1}

\usepackage{epsfig}
\usepackage{graphics,graphicx}

\usepackage{amssymb}
\usepackage{mathrsfs}
\usepackage{amsmath}
\usepackage{verbatim}
\usepackage{hyperref}

\def\be{\begin{equation}}
\def\ee{\end{equation}}

\begin{document}

\title{Effect of pressure on the anomalous response functions of  a confined
  water monolayer at low temperature}


\author{Marco G. Mazza}
\affiliation{Center for Polymer Studies and Department of Physics,  Boston University, Boston, Massachusetts 02215, USA}
\affiliation{Max Planck Institute for Dynamics and Self-Organization, D-37077 G\"ottingen, Germany}
\author{Kevin Stokely}
\affiliation{Center for Polymer Studies and Department of Physics,  Boston University, Boston, Massachusetts 02215, USA}
\author{H. Eugene Stanley}
\affiliation{Center for Polymer Studies and Department of Physics,  Boston University, Boston, Massachusetts 02215, USA}
\author{Giancarlo Franzese}
\affiliation{Departament de Fisica Fonamental, Universitat de Barcelona, Diagonal 647, 08028 Barcelona, Spain}

\begin{abstract}

We study a coarse-grained model for a water monolayer that cannot
crystallize due to the presence of confining interfaces, such as 
 protein powders or inorganic surfaces.  Using both Monte Carlo
simulations and mean field calculations, we calculate three response
functions: the isobaric specific heat $C_P$, the isothermal
compressibility $K_T$, and the isobaric thermal expansivity $\alpha_P$.
At low temperature $T$, we find two distinct maxima in $C_P$,
$K_T$ and  $|\alpha_P|$, all converging toward a
liquid-liquid critical point (LLCP) with increasing pressure $P$.  We
show that the maximum in $C_P$ at higher $T$ is due to the fluctuations
of hydrogen (H) bond formation and that the second maximum at lower $T$
is due to the cooperativity among the H bonds.  We discuss a similar
effect in $K_T$ and  $|\alpha_P|$. If this cooperativity
were not taken into account, both the lower-$T$ maximum and the LLCP
would disappear.  However, comparison with recent experiments on water
hydrating protein powders 
provides evidence for
the existence of the lower-$T$ maximum, supporting the hypothesized LLCP
at positive $P$ and finite $T$.  The model also predicts that when $P$
moves closer to the critical $P$ the $C_P$ maxima move closer in $T$
until they merge at the LLCP.  
Considering that other scenarios for water are thermodynamically possible,
we discuss how an experimental
measurement of the changing separation in $T$ between the two maxima of
$C_P$ as $P$ increases could determine the best scenario for describing
water.

\end{abstract}


\maketitle

\section{Introduction}

Because of their relevance to physics, chemistry, and biology, the
anomalies of water have attracted intense interest
\cite{Debenedetti-JPCM03,Angell-Sci08,Ball2011,BIF2012}.  One of water's
anomalies is its large isobaric specific heat $C_P$, which
\emph{increases} upon cooling below $35^\circ$C \cite{Franks}.  Two
other response functions, the isothermal compressibility $K_T$, and the
isobaric thermal expansivity $\alpha_P$ also increase in magnitude upon
cooling for a wide range of temperature $T$.  This increase is rapid in
the supercooled region, with a possible divergence between $T \approx
-48^\circ$C \cite{Angell-divergence} and $T \approx
-51^\circ$C \cite{Tombari1999}.  However, experimental data for
the bulk liquid state are only available down to $T_{\rm H} \approx
-41^\circ$C, due to homogeneous nucleation of ice.

Several different thermodynamic scenarios have been proposed to explain
the behavior of the response functions:

\begin{itemize}

\item[{(i)}] In the first, namely the \emph{stability limit\/} (SL) scenario
  \cite{speedy}, the liquid-gas spinodal in the negative pressure region
  bends upwards as $T$ decreases and reenters the positive pressure
  region at $T<T_{\rm H}(P)$. The liquid state is thus delimited by a
  single thermodynamic boundary $P_s(T)$. This scenario would explain
  the anomalous behavior of water because response functions diverge
  upon approaching a spinodal.

\item[{(ii)}] In the second, namely the \emph{singularity free\/} (SF) scenario
  \cite{Sastry-PRE96,Teixeira-JCP80}, the increase of the response
  functions upon cooling is a direct consequence of the negatively
  sloped locus of temperatures of maximum density (TMD) in the
  pressure-temperature ($P$--$T$) plane. No other thermodynamic cause is
  invoked.  In this scenario, $C_P$ reaches a finite maximum that does
  not change value with increasing $P$, but shifts to lower $T$, while
  $K_T$ and $|\alpha_P|$ have maxima that increase with increasing $P$
  and shift to lower $T$ \cite{Rebelo1998}.

\item[{(iii)}] A third, namely the \emph{liquid-liquid critical point} (LLCP)
  scenario, hypothesizes the existence of a first order liquid-liquid (LL)
  phase transition line,
  with negative slope in the $P$--$T$ plane, separating a low density
  liquid (LDL) and a high density liquid (HDL).  By moving along this
  line, the density difference between LDL and HDL decreases and
  disappears at a point $(P_c,T_c)$, which is the LLCP
  \cite{Poole-Nat92}.  The response functions diverge upon approaching the LLCP.  A
  locus of maxima of each thermodynamic response function emanates from
  the critical point into the one-phase region.  Since all thermodynamic
  response functions are proportional to the correlation length $\xi$,
  near the LLCP, each response function locus is well approximated by
  the \emph{Widom line}, defined to be the locus of maxima of $\xi$
  \cite{Limei-PNAS05,Franzese-JPCM07,Moore2009}.  Hence even in the one phase
  region at subcritical $P$, $C_P$ and the other response functions
  are expected to increase upon 
  approaching the Widom line.  Many studies place the LLCP at $P_c>0$
 \cite{Poole2005,Fuentevilla2006,Liu2009,Abascal2010,Poole2011,Kesselring2012,Holten}, 
  though some simulations suggest $P_c<0$ \cite{Tanaka-negllcp}.

\item[{(iv)}] The \emph{critical-point free} (CPF) scenario
  \cite{Angell-Sci08,Poole-PRL94} hypothesizes the presence of a
``order-disorder'' transition without a critical point.
  This transition may be first-order in nature.  Here the response
  functions increase
  upon cooling because they approach the ``order-disorder'' transition,
  or the limit of stability associated with a first order transition.
  This limit of stability fulfills the requirements of the
  stability-limit (SL) conjecture \cite{Stokely-PNAS10}.

\end{itemize} 

Although experiments on bulk water below $T_{\rm H}$ have not been
possible due to ice nucleation, several studies have been carried out at
colder $T$ in confined environments.  Under appropriate conditions,
confined geometries destroy the long-range order necessary for crystal
formation \cite{confinement}.  The relation of confined water to bulk
water is debated
\cite{Soper08,StrekalovaPRL2011,StrekalovaJPCM2012,StrekalovaPRL2012,XuMolinero2011}, 
but the 
behavior of confined water could provide insights into the behavior of
bulk water.  Confined water itself is of greater interest
because it is essential to a number of physical processes in geology
\cite{geophys}, meteorology \cite{meteo}, chemistry
\cite{MerteScience2012}, and biology \cite{israelach}.

The dynamics of the H bond network have been studied for water confined
to the surface of the globular protein lysozyme.  Some authors report a
crossover in the $T$-dependence of the relaxation time for H bond
reorientation \cite{Chen-PNAS06,Frauenfelder-PNAS09}.  A new analysis
reveals the existence of two distinct crossovers in the supercooled
regime \cite{Mazza-PNAS11}. Through direct calculations these results
have been related to a novel behavior of $C_P$, suggesting two separate
maxima \cite{Mazza-PNAS11}.

Here, we study a coarse-grained model for a monolayer of water. We use
Monte Carlo (MC) simulations and analytic mean field (MF) calculations
to determine how $C_P$, $K_T$, and $\alpha_P$ change for increasing $P$.
The model is general enough to be able to reproduce the four scenarios
described above \cite{Stokely-PNAS10} for different values of the model
parameters, as well as the experimental results for the dynamics and the
predicted behavior of $C_P$ \cite{Mazza-PNAS11}.

We find for all cases in which the model exhibits a LLCP or
LL phase transition that the response functions exhibit two maxima at low $T$ and
that the dynamics, as a consequence, has two distinct crossovers.  
As we will discuss in the following, the
recent measurements indicating that water hydrated protein powders 
exhibits two dynamic crossovers \cite{Mazza-PNAS11} rule out the SF
scenario as a realistic description of water. The temperatures at which
the two maxima of response functions are located depends on $P$.  For $P < P_c$, the
maxima move closer in $T$ with increasing $P$, while for $P > P_c$ the
maxima move further apart in $T$ with increasing $P$.  Based on previous studies
\cite{Stokely-PNAS10} we can conclude that the response functions
for the CPF scenario behave
as they do for $P>P_c$ in the LLCP scenario. Thus these findings suggest
that an experimental study of how the temperatures of the two 
maxima of $C_P$, $K_T$, and $\alpha_P$ depend on $P$ would be a test
for which scenario best describes water, as well as a method of
estimating $(P_c,T_c)$. 

The structure of the paper is as follows. In Sec.~\ref{model} we
describe the model used for the MC simulations and the MF
calculations. Section~\ref{results} reports our results, which are
discussed in Sec.~\ref{disc}. We present our conclusions in Sec.~V.

\section{Model}\label{model}

\subsection{Coarse Grained Model for Monte Carlo Calculations}

The system consists of $N$ particles in a 3-dimensional monolayer
occupying a volume $V$, which is divided into $N$ equivalent cells, each
with one molecule $i\in[1,N]$ and with volume $V/N$ larger than a
hard-core volume $v_0$.  We model water molecules in a confined
environment, which have fewer nearest-neighbor (n.n.) molecules than
bulk water \cite{koga}.  Because of the confinement, and to keep our
model simple, we fix the number of n.n. to four, consistent with
atomistic simulations of a water monolayer between confining walls
separated by $\approx 0.5$~nm \cite{Zangi,Kumar}.  By coarse-graining
the position of water molecules within each cell, we reduce our
representation of the monolayer to a 2-dimensional system, partitioned
into square cells, which preserves the number of n.n. molecules.

The interaction Hamiltonian is
\cite{Franzese-JPCM02,PhysA02,Franzese-PRE03,KumarPRL2008,KumarJPCM2008,JPCM08,JPCM09,JPCM2010,FBS2011,JPCB2011,PRER2012} 
\begin{equation}
\mathscr{H} \equiv -J \displaystyle \sum_{\langle i,j\rangle} n_in_j
\delta_{\sigma_{ij},\sigma_{ji}} - J_\sigma \displaystyle \sum_i n_i
\sum_{(k,l)_i}\delta_{\sigma_{ik},\sigma_{il}} + U_W(r), 
\label{EQ1}
\end{equation}
where to each cell we associate a variable $n_{i}=0,1$.  If cell $i$ has
a density $\rho_{i}>\rho_0/2$, with $\rho_0=1/v_0$ and
$\rho_i/\rho_0\leq 1$, then the cell is liquid-like and $n_{i}=1$. If
$\rho_{i}\leq \rho_0/2$, then the cell is gas-like and $n_{i}=0$.

The first term in Eq.~(\ref{EQ1}) represents the covalent (directional)
H bond component, where $J>0$ represents the covalent energy gained per
H bond.  Here, $\sigma_{ij}=1,\ldots,q$ are Potts variables representing
the bond indices of molecule $i$ with respect to its n.n. molecules $j$,
and $\langle i,j \rangle$ denotes that $i$ and $j$ are n.n.  We choose
the parameter $q$ by selecting $30^\circ$ as the maximum deviation from
linear bond (i.e., $q = 180^\circ/30^\circ = 6$). Hence, every molecule
has $q^4 = 1,296$ possible configurations.  A H bond is formed between
two n.n. molecules $i$ and $j$ if and only if both are in
liquid-like cells ($n_{i}n_{j}=1$) and their variables $\sigma_{ij}$ and
$\sigma_{ji}$ are in the same state
($\delta_{\sigma_{ij}\sigma_{ji}}=1$, with $\delta_{ab}=1$ if $a=b$, 0
otherwise). The first condition specifies $r_0\leq r < \sqrt{2} r_0$,
with $r_0\equiv \sqrt{v_0/h}$, $h$ being the monolayer thickness, and
the second condition specifies that both molecules must have the correct
relative orientation to form a H bond. Thus the use of the bonding
variables $\sigma_{ij}$ allows us to take into account not only the
decrease of energy, but also the decrease of orientational entropy due
to the formation of H bonds.

The second term in Eq.~(\ref{EQ1}) accounts for the many-body
interaction/cooperative effect that characterizes water \cite{Frank57}
and it has an intrinsically quantum nature \cite{ludwig2001}. This
interaction is responsible for the O--O--O correlation in bulk water
\cite{Ricci}, locally driving the molecules toward a tetrahedral
configuration. A pair of bond indices of a molecule in the same state
corresponds to a minimization of the many-body interaction, the energy
is decreased by an amount $J_\sigma\geqslant0$ per pair of such indices,
and $(k,l)_i$ indicates the set of six different pairs of the four bond
indices of molecule $i$.

The third term $U_W(r)$ denotes the isotropic component of the
water-water interaction due to van der Waals dispersion forces and
short-range repulsion and is represented by a modified Lennard-Jones
potential between molecules at distance
\begin{equation}
r\equiv(V_0/Nh)^{1/2}, 
\label{r}
\end{equation}
where $V_0$ is defined below, with attractive energy $\epsilon>J$ and
with a hard-core repulsion
\begin{equation}
U_W(r)\equiv\begin{cases}
\infty & \text{if $r\leqslant r_0$,}\\
\epsilon\left[\left(\frac{r_0}{r}\right)^{12}-
      \left(\frac{r_0}{r}\right)^6\right] & \text{if $r>r_0$.}\\
\end{cases}
\label{U}
\end{equation}
Note that, by the definition used here, the minimum of $U_W(r)$ is
$\epsilon/4$.

Experiments show that liquid water has a tendency to acquire a local
tetrahedral order in the bulk at low $T$ and low $P$ up to the second
shell, due to the formation of an average of four H bonds per molecule
\cite{SoperRicci2000}. By increasing $T$ or $P$ the H bond network is
partially disrupted leading to the formation of a more compact local
structure characterized by a less clear separation between the first and
second shell \cite{SoperRicci2000}, by a larger coordination number due
to a molecule of the second shell moving toward the first shell in an
interstitial position \cite{Saitta2003}, and by a larger local density,
i.e., by a smaller volume per molecule on average. We take into account
this volume effect associated with the formation and breaking of H bonds
by assuming that the total volume is
\begin{equation}
V\equiv V_0+N_{\rm  HB}v_{\rm HB},
\label{V}
\end{equation}
where $V_0\geqslant N v_0$ is a dynamic variable that fluctuates in the
simulations and corresponds to the volume of the system without H bonds,
$v_{\rm HB}$ is the average volume increase per H bond that results from
the difference between the high-density local structure and the
low-density local structure found in the experiments, and
\begin{equation} 
N_{\rm HB}\equiv\sum_{\langle i,j\rangle}n_in_j\delta_{\sigma_{ij},\sigma_{ji}}
\end{equation}
is the total number of H bonds.

Note that, by the definition of $r$ in Eq.~(\ref{r}), the increase of
volume per molecule associated with the formation of a H bond does not
affect the calculation of the isotropic interaction in
Eq.~(\ref{U}). This choice reflects the experimental finding that the
decrease of local density due to the formation of the H bonded
tetrahedral structure does not affect the average water-water distance,
but only second-neighbor distances \cite{SoperRicci2000}.

We perform MC simulations for $N=10^4$ molecules at constant $P$ and
$T$.  The MC dynamics consists in updating the variables $\sigma_{ij}$
by means of the Wolff algorithm \cite{Wolff,mazzaCPC}, based on an
appropriate percolation approach
\cite{CataudellaPRL94,CataudellaPRE96,FranzesePRE00}.
The Wolf algorithm allows us to simulate the system in efficient way, with short
correlation times even at very low $T$ \cite{mazzaCPC}.
We update the
volume $V_0$ in accordance with the acceptance probability
$\min\left(1,\exp\left[-\beta\left(\Delta E+P\Delta V-Nk_{\rm
    B}T\ln(V_f/V_i)\right)\right]\right)$.  Here $\Delta E$ is the
variation of the right hand side of Eq.(\ref{EQ1}) with the update,
$\beta\equiv(k_{\rm B}T)^{-1}$, $k_{\rm B}$ is the Boltzmann constant, $\Delta
V\equiv V_f-V_i$ where $V_i$ and $V_f$ are the initial and final values
of the volume in Eq.~(\ref{V}), respectively.

As a consequence of our definition of $r$ in Eq.~(\ref{r}), $n_i=n_j$
and $\rho_i=\rho_j=\rho_0\equiv V_0/N$ for any $i$ and $j$, $\rho_0$
being the total density irrespective of the local density variation due
to H bonds. Thus, at the coexistence between two phases, e.g., the
liquid and the gas, the cells flip their state between liquid-like
($n_i=1$) and gas-like ($n_i=0$) together, the whole system being
homogeneous.

\subsection{Coarse Grained Model for Mean Field Calculations}

The homogeneity condition, which we adopt in the MC simulations, is no
longer necessary when we solve the model within a MF approximation where
each cell $i$ has an a priori different number density $n_i = 0,1$.  The
Hamiltonian in this case is
\begin{equation} 
\mathscr{H} \equiv -J \sum_{\langle i,j\rangle}
n_in_j\delta_{\sigma_{ij},\sigma_{ji}} - J_\sigma \sum_i
n_i \sum_{(k,l)_i}\delta_{\sigma_{ik},\sigma_{il}} - \epsilon
\sum_{\langle i,j\rangle} n_i n_j . 
\label{MF}
\end{equation}
The qualitative behavior of the model remains similar, though the
liquid-gas critical point $C$ is moved to lower $T$ and $P$. It can be
shown \cite{oriol} that the discrepancy in the estimate of the
parameters of the liquid-gas critical point $C$ between the MF and the
MC calculations is primarily due to the homogeneity condition imposed in
the MC case, as discussed above, which makes the liquid phase more stable
than the gas phase.

Here, for both MC and MF calculations, we study the model for parameters
$J/\epsilon = 0.5$, $J_\sigma/\epsilon = 0.05$, and $v_{\rm HB}/v_0 =
0.5$.  This choice of parameters is discussed in
Ref.~\cite{Stokely-PNAS10} and has proven to be comparable to the
experiments \cite{Mazza-PNAS11}.

In the following, all $T$ are reported in units of $\epsilon/k_{\rm B}$,
and $P$ in units of $\epsilon/v_0$.  For this choice of parameters the
model exhibits in the MC simulations a LLCP at $P_c = 0.70 \pm 0.1$ and $T_c = 0.05\pm
0.01$.  For $P>P_c$ there exists a first-order LL phase
transition with negative slope in the $P$--$T$ plane.  We study
pressures in the interval 0.001 $\leqslant P \leqslant$ 1.5.

\section{Results}\label{results}

\subsection{Isobaric Specific Heat}

We calculate the isobaric specific heat
\begin{equation}
C_P\equiv\left(\partial H/\partial T\right)_P,
\label{Cp}
\end{equation}
where 
\begin{equation}
H\equiv \langle \mathscr{H}  \rangle+P\langle V \rangle
\label{H}
\end{equation}
is the enthalpy, and $\langle \cdot \rangle$ denotes the thermodynamic
average.

\subsubsection{Monte Carlo Calculations}

From our MC simulations, we find that for low $P$ isobars, such as
$P=0.001$, the model exhibits two $C_P$ maxima in the liquid state
\cite{Mazza-PNAS11}.  The maximum at higher $T$ is broad, while the
maximum at lower $T$ is rather sharp (Fig.~\ref{MC-CP}a).

Using MC simulations we find for several $P < P_c$ that the temperatures
of the maxima of $C_P$ depend on $P$.  As $P$ increases toward $P_c$,
the sharp maximum remains relatively constant in $T$, while the
higher-$T$ broad maximum moves to lower $T$.  For $P \approx P_c$, the
two maxima merge.  The value of the sharp maximum slowly increases with
increasing $P$, reaching the largest values at $P_c$ \cite{Note1}.

When $P>P_c$ the sharp maximum at lower-$T$ occurs at the temperature of
the first-order LL phase transition (Fig.~\ref{MC-CP}b).
As $P$ increases far above $P_c$, the two maxima again separate in $T$.
The sharp maximum decreases in value, and moves to lower $T$ with
increasing $P$, following the LL phase transition.  The broader $C_P$
maximum at higher $T$ becomes independent of $P$, as has been noted
\cite{Marques-PRE07,Note2}.  Hence as $P$ continues to increase, the
maxima become further separated in $T$.

\subsubsection{Mean Field Calculations}

We also calculate $C_P$ within a MF approximation
\cite{Franzese-JPCM02,PhysA02,Franzese-PRE03,KumarPRL2008,KumarJPCM2008,JPCM08,JPCM2010,PRER2012}.
For $P<P_c$ we find qualitative behavior similar to that found in MC
simulations (Fig.~\ref{MF-CP}a).  For $P=0$, $C_P$ exhibits two maxima.
Both maxima move to lower $T$ as $P$ increases, though the broader
maximum has a $P$-dependence that is more pronounced than that found in
MC simulations.  In MF, the two maxima are distinct only significantly
below $P_c$; above $P \simeq 0.3$ both peaks merge into a single
maximum.  The $C_P$ maximum increases on approaching the MF critical
pressure $P_c^{\rm MF} = 0.82 \pm 0.04$ (Fig.~\ref{MF-CP}b).  For
$P>P_c^{\rm MF}$, $C_P$ exhibits only one maximum, marking the LL phase
transition line.  The higher-$T$ maximum at $P>P_c$ is not seen in the
MF treatment of the model, as it is likely that bond variables 
satisfying  the directional bond interaction and the cooperative bond
interaction are not independent.

\subsection{Origin of the two maxima in $C_P$}

The origin of the two distinct maxima in $C_P$ may be understood by
considering the enthalpy to be a sum of a contribution due to single H
bond formation ($H^{\rm 1HB}$) and a term due to the cooperative
interaction among bonds ($H^{\rm coop}$), i.e.,
\begin{equation}
H = H^{\rm 1HB} + H^{\rm coop},
\label{H2}
\end{equation}
with
\begin{equation}
\begin{aligned}
H^{\rm 1HB} &\equiv \langle -J N_{\rm HB}+ PN_{\rm HB}v_{\rm HB}\rangle\\
H^{\rm coop} &\equiv H-H^{\rm 1HB} .
\end{aligned}
\end{equation}
Here, $H^{\rm 1HB}$ contains all terms proportional to $N_{\rm HB}$, and
$H^{\rm coop}$ includes the enthalpy of the cooperative interaction, as
well as the contribution coming from the van der Waals interaction,
which is negligible in the range of $T$ of interest here.

From Eq.~(\ref{H2}) we derive
\begin{equation}
C_P=C_P^{\rm 1HB}+C_P^{\rm coop},
\label{Cp2}
\end{equation}
where, by definition, for the MC model
\begin{equation}
\begin{aligned}
C_P^{\rm 1HB} \equiv\left(\partial H^{\rm 1HB}/\partial T\right)_P &= -
(J -P v_{\rm HB}) (\partial \langle N_{\rm HB} \rangle /\partial
T)_P,\\ 
C_P^{\rm coop} \equiv 
\left(\partial H^{\rm coop}/\partial T\right)_P &= P\left(\partial V_0/\partial T\right)_P
- J_\sigma (\partial \langle N_{\rm coop} \rangle /\partial T)_P +
(\partial \langle U_W(r) \rangle/\partial T)_P,
\end{aligned}
\label{Cp2-terms}
\end{equation}
and
\begin{equation}
N_{\rm coop} \equiv \sum_i
n_i \sum_{(k,l)_i}\delta_{\sigma_{ik},\sigma_{il}}
\label{Ncoop}
\end{equation}
is the total number of bond-index pairs that on each molecule minimize
the cooperative interaction.  In the low-$T$ region that we explore in
this work, the isobaric variation of $V_0$ and $\langle U_W(r) \rangle$
with $T$ is negligible. Therefore, for the liquid at $T$ far below the
liquid-gas transition, we can write
\begin{equation}
C_P^{\rm coop} \approx - J_\sigma (\partial \langle N_{\rm coop} \rangle /\partial T)_P.
\label{approx}
\end{equation}

A similar decomposition can be written also for the MF model, by
replacing $\langle U_W(r) \rangle$ with $-\langle \epsilon
\sum_{\langle i,j\rangle} n_i n_j \rangle$ in Eq.~(\ref{Cp2-terms})
and observing that its isobaric variation with $T$ is negligible in
the low-$T$ region studied here.

\subsubsection{Monte Carlo Calculations}

\begin{figure}
\includegraphics[scale=0.5]{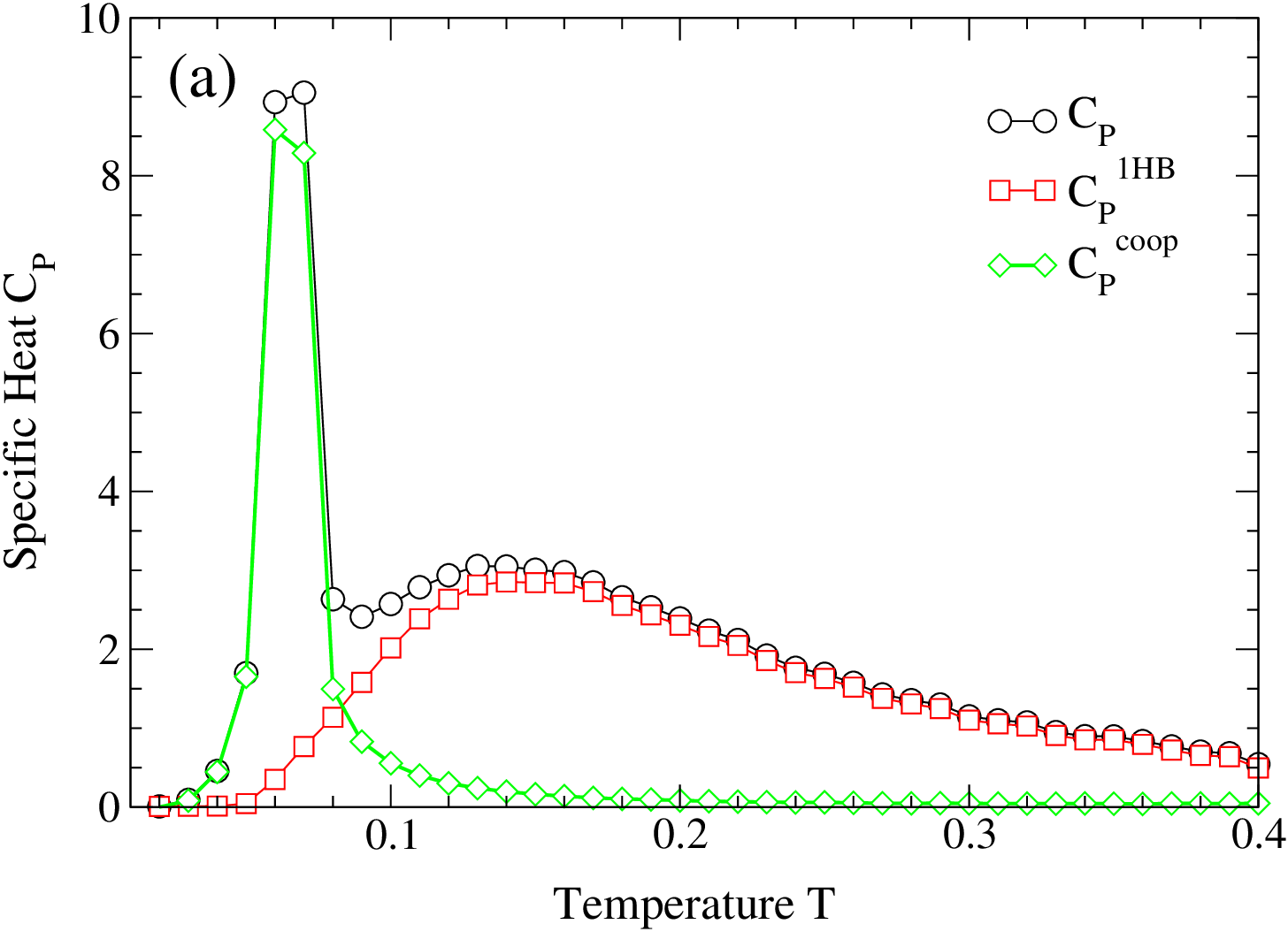}
\includegraphics[scale=0.5]{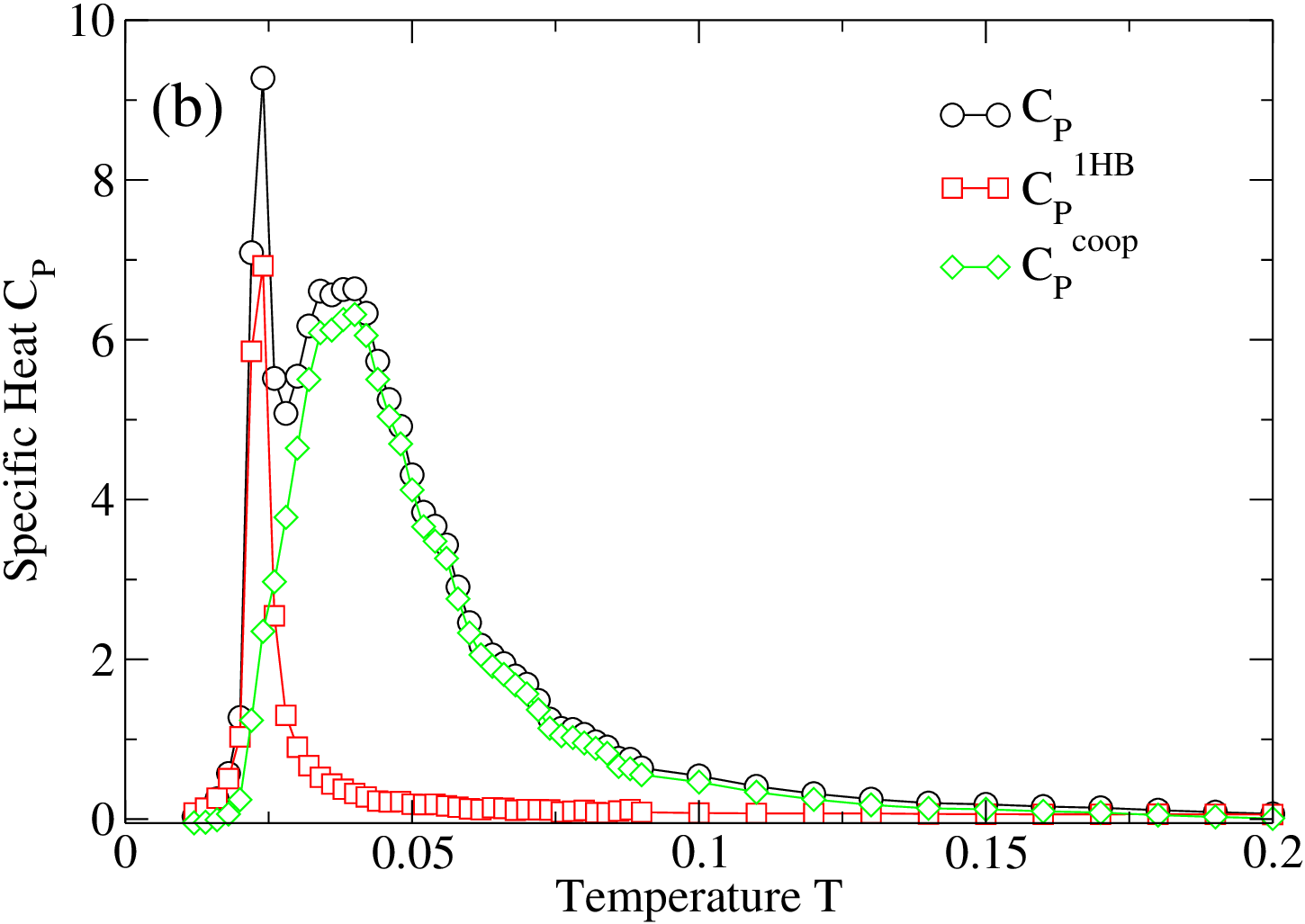}
\caption{{\it Monte Carlo Calculations:\/} Decomposition of $C_P$ into
  the components $C_P^{\rm coop}$ and $C_P^{\rm 1HB}$, as in
  Eqs.~(\ref{Cp2}-\ref{approx}), (a) for $P=0.1$, and (b) for $P=0.94$.
  Note that at low $P$ the high-$T$ broad $C_P$ maximum is due to
  $C_P^{\rm 1HB}$ and the sharp maximum at low $T$ is due to $C_P^{\rm
    coop}$. Vice versa at high $P$ the broader maximum at high $T$ is due to
  $C_P^{\rm coop}$ and the sharp maximum at low $T$ to $C_P^{\rm 1HB}$,
  inverting the order.}
\label{DECOMP}
\end{figure}

To understand which term in Eq.~(\ref{Cp2}) is responsible for each
maximum in $C_P$, we calculate separately the two contributions, as in
Eqs.~(\ref{Cp2-terms})--(\ref{approx}), and compare them with the direct
calculation of $C_P$ from Eq.~(\ref{Cp}). We find that each term
accounts for one and only one of the two maxima of $C_P$, as shown in
Fig.~\ref{DECOMP}. Furthermore, we observe that, for $P<P_C$
(Fig.~\ref{DECOMP}a), $C_P^{\rm 1HB}$ is responsible for the maximum at
higher $T$, while $C_P^{\rm coop}$ is responsible for the maximum at
lower $T$. On the other hand, for $P>P_C$ (Fig.~\ref{DECOMP}b), the two
maxima invert their order, with the high-$T$ maximum due to $C_P^{\rm
  coop}$ and the low-$T$ maximum to $C_P^{\rm 1HB}$, and interchange
their shape,  with the one due to $C_P^{\rm 1HB}$ becoming broader and
the one due to $C_P^{\rm coop}$ becoming sharper.

Equations~(\ref{Cp2-terms}--\ref{approx}) give us the key to
understanding the nature of these two $C_P$ maxima.  By definition $C_P$
is proportional to the isobaric variation of entropy with $T$. Therefore
each maximum in $C_P$ corresponds to a maximum in the change of entropy,
i.e., a maximum in a structural change. In particular,
Eqs.~(\ref{Cp2-terms})--(\ref{approx}) emphasize

\begin{itemize}

\item[i)] that the maximum in $C_P^{\rm 1HB}$ is associated with the
  largest isobaric variation of the number $N_{\rm HB}$ of H bonds with
  $T$, and

\item[ii)] that the maximum in $C_P^{\rm coop}$ is due to the largest
  variation with $T$ of the number $N_{\rm coop}$ of H bonds that
  minimize the cooperative interaction at constant $P$.

\end{itemize}

These correspondences are verified by direct calculations of $(\partial
\langle N_{\rm HB} \rangle /\partial T)_P$ and $(\partial \langle N_{\rm
  coop} \rangle /\partial T)_P$ (Figs.~\ref{MC-NHB} and \ref{MC-NIM}).
We find that the loci of state points at which these derivatives are at
a maximum overlap with the loci of maxima of $C_P^{\rm 1HB}$ and
$C_P^{\rm coop}$, respectively.  In Sec.~\ref{disc} we will discuss in
more details the
physical interpretation of these results.

\subsubsection{Mean Field Calculations}

\begin{figure}
\includegraphics[scale=0.5]{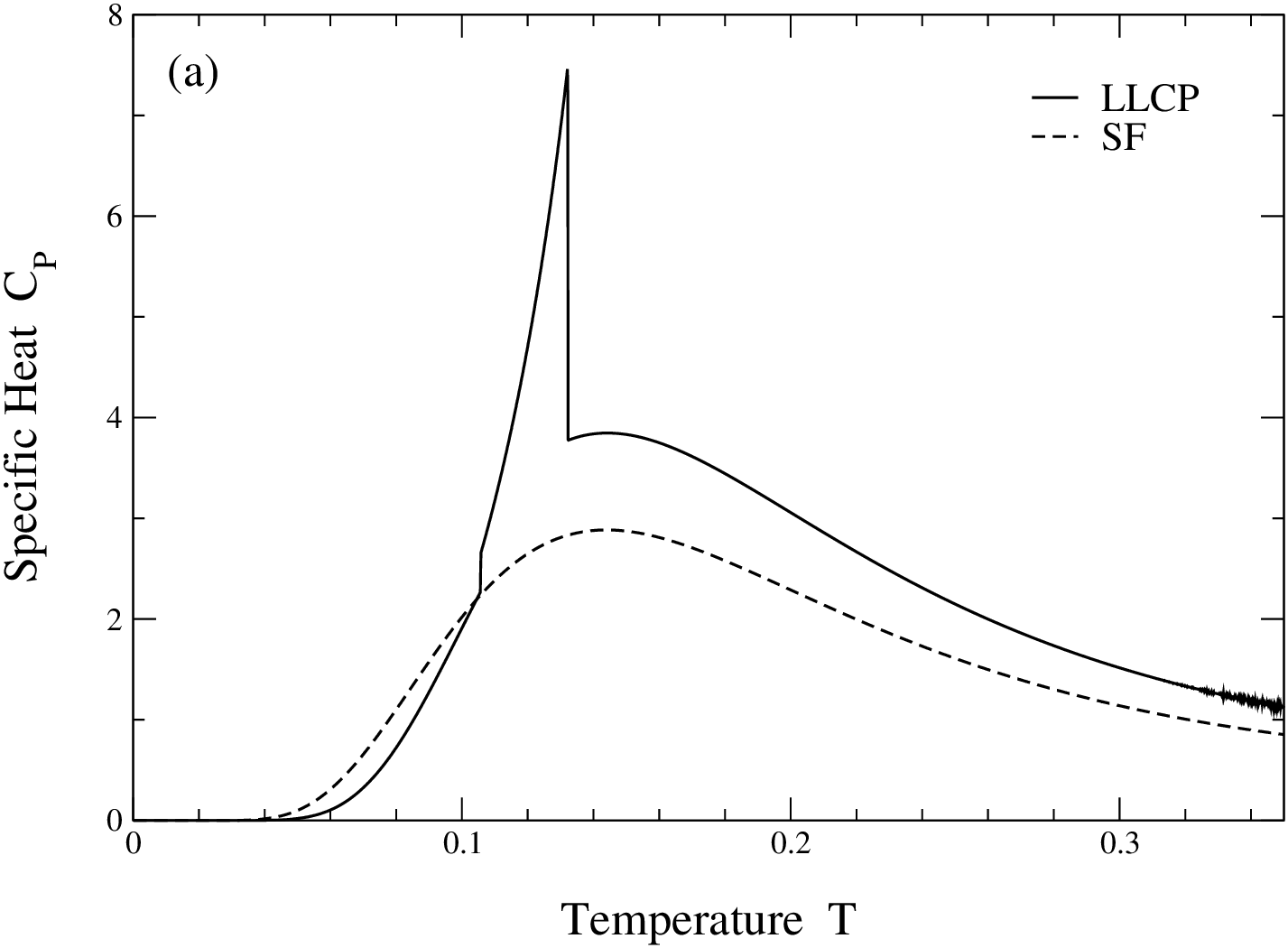}
\includegraphics[scale=0.5]{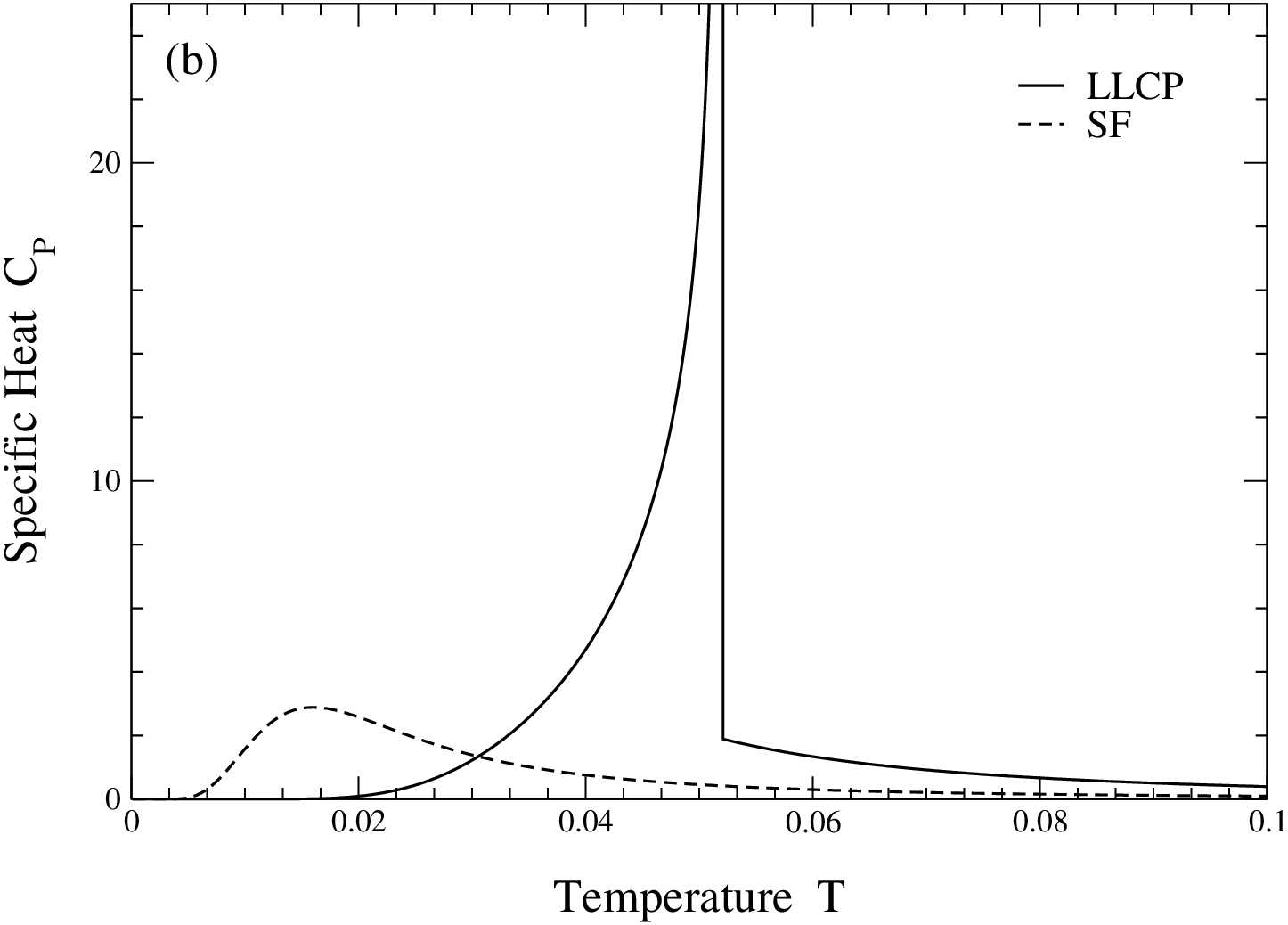}
\caption{{\it Mean Field Calculations:\/} Comparison of $C_P$
  calculations for the LLCP scenario case ($J_\sigma/\epsilon=0.05$) and
  the SF case ($J_\sigma=0$) (a) for $P=0.1$, and (b) for $P=0.9$.  At
  low $P$, (a), the low-$T$ maximum is present only in the LLCP case,
  indicating that it is due to the cooperative term with $J_\sigma\neq
  0$ in Eq.~(\ref{MF}).  At high $P$, (b), due to the mean field approximation
  we use, we find in both cases one single maximum in $C_P$.  In the
  LLCP case, the maximum occurs at the LL phase transition temperature,
  while it occurs at lower $T$ in the SF case, for which no LL phase
  transition occurs.
\label{DECOMP-MF}}
\end{figure}

Although the decomposition in Eq.~(\ref{Cp2}) also applies to the MF
model, in our MF approximation to solve the model, described in detail
in Refs.~\cite{Franzese-JPCM07,Stokely-book}, we obtain
\begin{equation}
C_P \equiv
\left(\frac{\partial H}{\partial T}\right)_P 
\approx 
2 (J-Pv_{\rm
HB}+3J_\sigma) \left(\frac{\partial p_\sigma}{\partial T}\right)_P,
\label{CpMF}
\end{equation}
where $p_\sigma$ is the probability that the facing bonding variables of
two nearest neighbor molecules will be in the same state, but not
necessarily in the state that minimizes the $J_\sigma$ cooperative
interaction. The approximation consists in neglecting 
the liquid-gas contribution at low $T$.  Because this expression for the
specific heat cannot be easily separated into two terms, we calculate
$C_P$ for the model either with or without cooperative interactions
(Fig.~\ref{DECOMP-MF}).  When including cooperative interactions
($J_\sigma>0$) the LLCP is present. Without them ($J_\sigma=0$) the SF
scenario is obtained.

\begin{figure}
\includegraphics[scale=0.5]{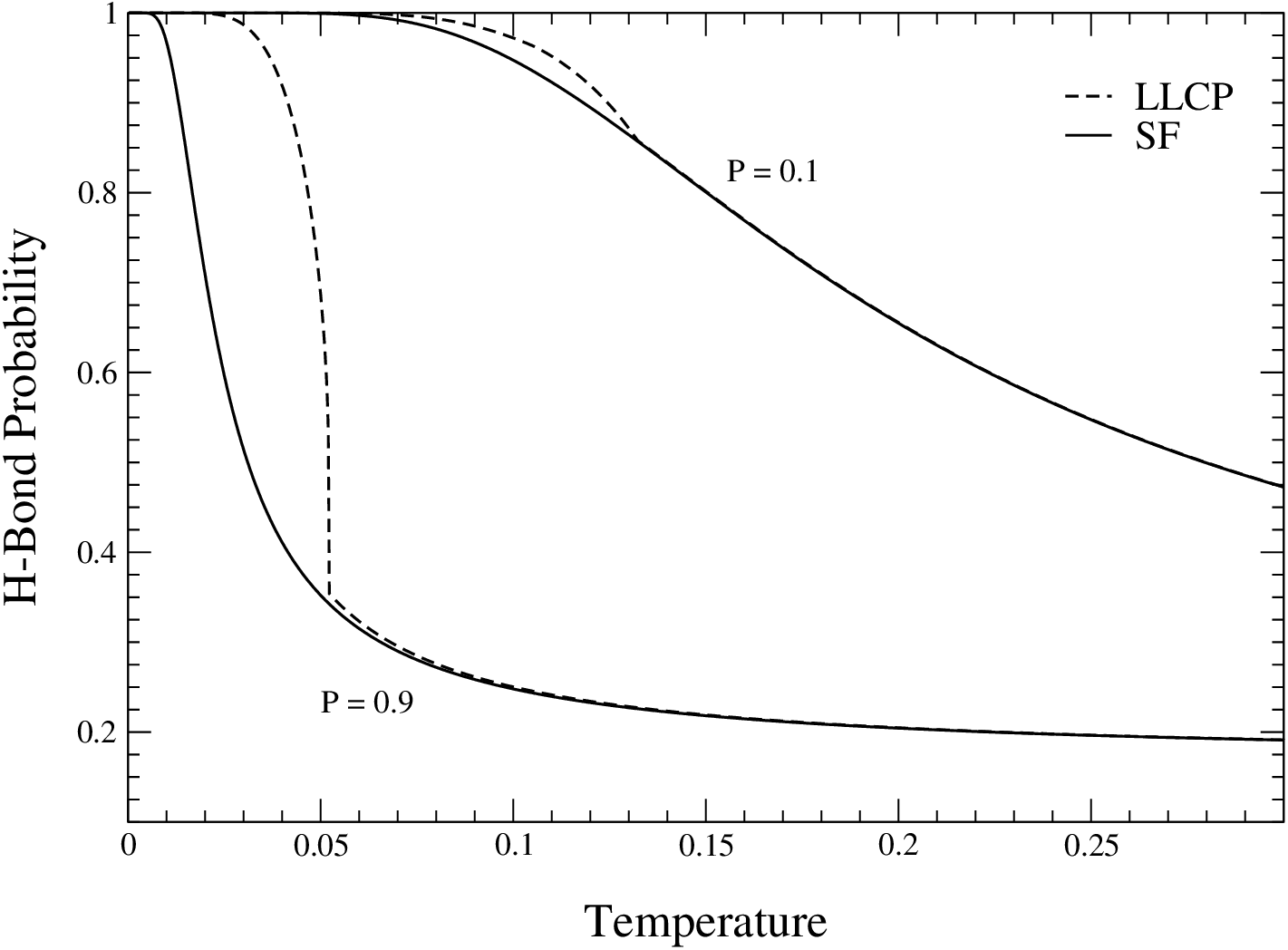}
\caption{{\it Mean Field Calculations:\/} Probability of forming a
  H bond in the mean field approximation as a function of $T$, from which we
  calculate $C_P$, for $P=0.1.$ (upper curves) and $P=0.9.$ (lower
  curves), as indicated by the labels.  The calculations for the LLCP
  scenario case ($J_\sigma/\epsilon=0.05$) are shown as dashed lines,
  and those for the SF case ($J_\sigma=0$) as continuous lines.
\label{NHB-MF}}
\end{figure}

In the SF scenario the model is exactly solvable. We find that $C_P$
shows only one maximum, which is related to the isobaric $T$-derivative
of $\langle N_{\rm HB}\rangle$ \cite{Sastry-PRE96} 
(Fig.~\ref{NHB-MF}).  We find that at low $P$ (Fig.~\ref{DECOMP-MF}a) the
SF maximum reproduces the LLCP high-$T$ broad maximum, has the same
shape, and occurs at the same $T$.  Thus the MF broad $C_P$ maximum at
low $P$ is not related to the cooperative interaction, proportional to
$J_\sigma$. Instead, because the sharper maximum at low $P$ is present
only in the LLCP scenario, we conclude that it is due to the effect of
the $J_\sigma$ cooperative term on the probability $p_{\rm HB}\equiv
N_{\rm HB}/4N$ (Fig.~\ref{NHB-MF}) as a consequence of the maximum
variation with $T$ of $\langle N_{\rm coop}\rangle$ at constant $P$. We
thus find in MF indirect evidence that validates the proposed mechanism
based on our MC calculations.

At $P>P_c$ (Fig.~\ref{DECOMP-MF}b), we observe in our MF solution only
one maximum in $C_P$ with two large asymmetric tails, instead of the two
maxima found in our MC calculations (Fig.~\ref{DECOMP}b). We understand
that this difference is caused by our MF approximation,
Eq.~(\ref{CpMF}). Nevertheless, we compare the calculations of $C_P$
without the $J_\sigma$ term to those with $J_\sigma$ term. We observe
that without the $J_\sigma$ term (SF case), $C_P$ has a broad maximum at
$T$ that is lower than the maximum found when the $J_\sigma$ term is
present (LLCP case). This difference is evident from the behavior of
$p_{\rm HB}$ in the two cases (Fig.~\ref{NHB-MF}).  Thus at $P>P_c$ the
cooperative $J_\sigma$ interaction contributes to $C_P$ at a $T$ that is
higher than the contribution coming from the non-cooperative term,
consistent with what we find in our MC calculations.

\subsection{Isothermal Compressibility and Isobaric
Thermal Expansivity}

We also calculate the isothermal compressibility $K_T$ and the isobaric
thermal expansivity $\alpha_P$, also known to exhibit anomalous behavior
in bulk water.  As with $C_P$, each of these depends upon $\langle
N_{\rm HB}\rangle$
\begin{align}
K_T &= \frac{1}{V_0 + \langle N_{\rm HB}\rangle v_{\rm HB}} \left[ \left(\frac{\partial
V_0}{\partial P}\right)_T + v_{\rm HB}\left(\frac{\partial \langle N_{\rm HB}\rangle}{\partial
P}\right)_T \right] \approx \frac{v_{\rm HB}}{V_0 + \langle N_{\rm HB}\rangle v_{\rm HB}}
\left(\frac{\partial \langle N_{\rm HB}\rangle}{\partial P}\right)_T , 
\label{MF-K}\\
\alpha_P &= \frac{1}{V_0 + \langle N_{\rm HB}\rangle v_{\rm HB}} \left[ \left(\frac{\partial
V_0}{\partial T}\right)_P + v_{\rm HB}\left(\frac{\partial \langle N_{\rm HB}\rangle}{\partial
T}\right)_P \right] \approx \frac{v_{\rm HB}}{V_0 + \langle N_{\rm HB}\rangle v_{\rm HB}}
\left(\frac{\partial \langle N_{\rm HB}\rangle}{\partial T}\right)_P ,
\label{MF-a}
\end{align}
where we use the observation that at low $T$ the variation of $V_0$  with $P$ and $T$ is
negligible.
 
\subsubsection{Monte Carlo Calculations}

\begin{figure}
\includegraphics[scale=0.5]{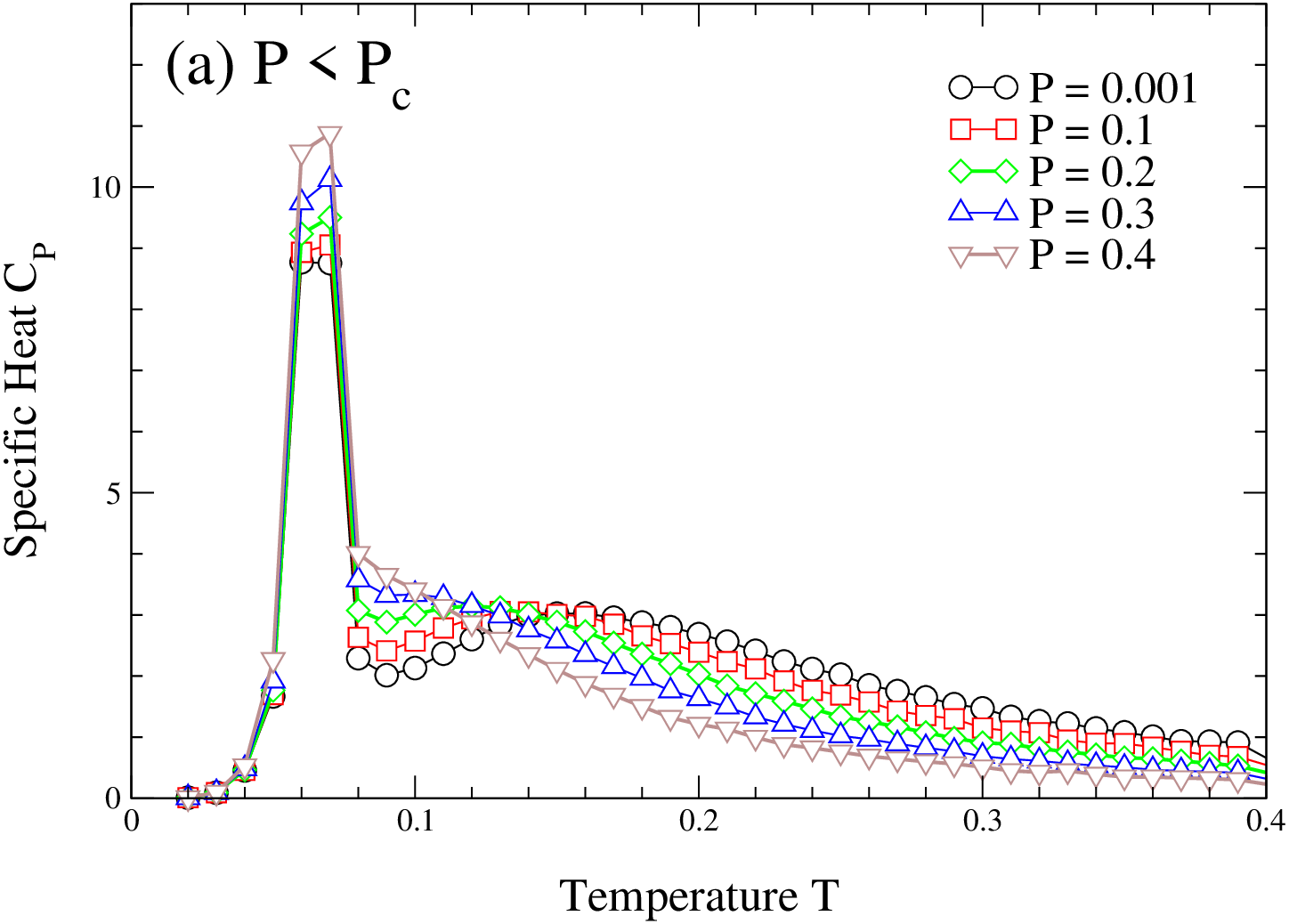}
\includegraphics[scale=0.5]{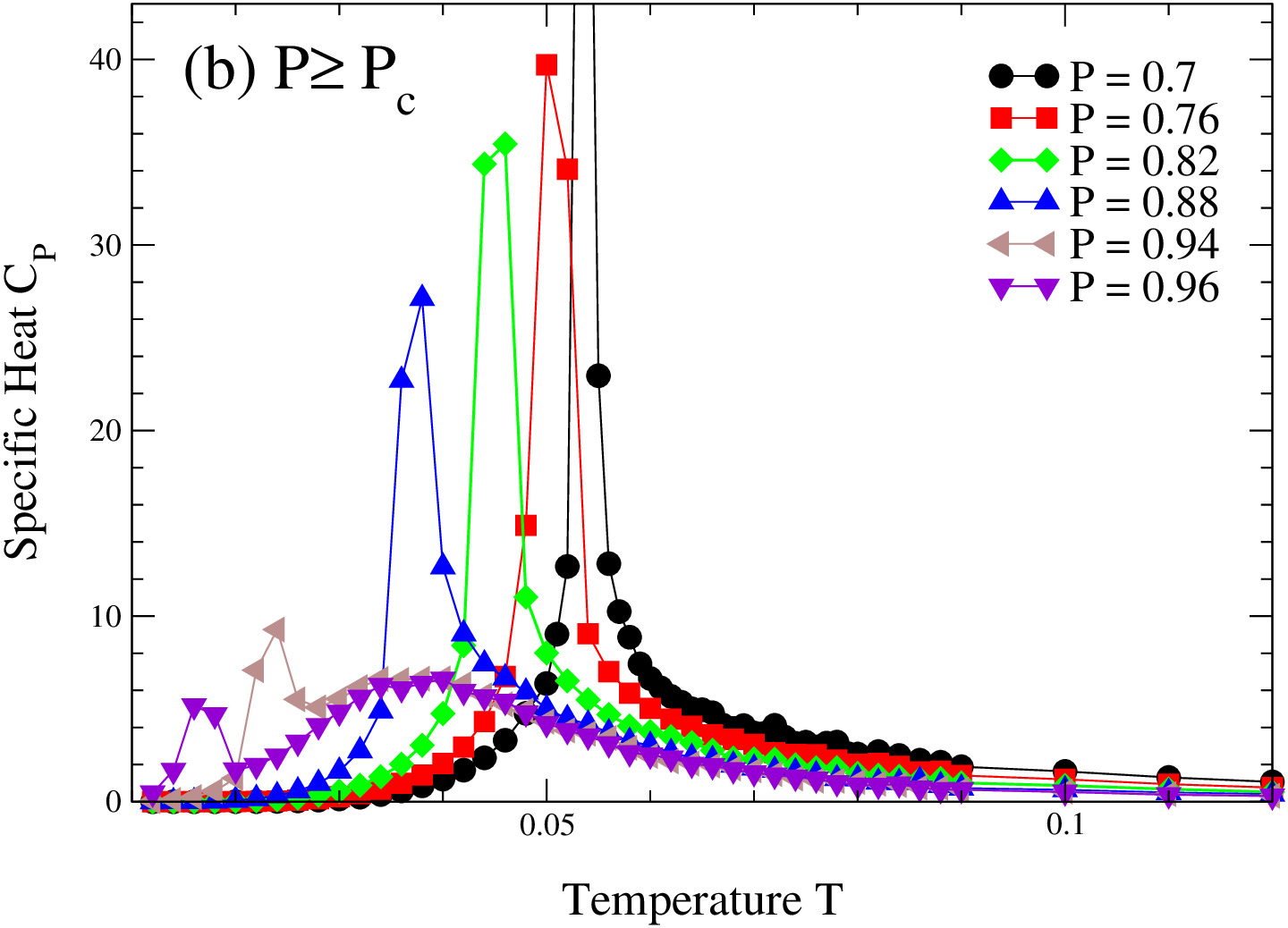}
\caption{{\it Monte Carlo Calculations:\/} (a) Temperature dependence of
  the specific heat $C_P$, for the parameters in the
  text, along low pressure isobars with $P<P_c$. A broad maximum is
  visible along with a more pronounced one at lower $T$. The first
  maximum moves to lower $T$ as the pressure is raised and it merges
  with the low--$T$ maximum at $P \approx 0.4$. Upon approaching $P_c =
  0.70 \pm 0.1$, the sharp maximum increases in value. (b) Same for $P
  \geq P_c$: the two maxima are separated only for $P > 0.88$; the sharp
  maximum decreases as $P$ increases. In both panels errors are smaller
  than symbol size and lines are guides for the eyes.}
\label{MC-CP}
\end{figure}

\begin{figure}
\includegraphics[scale=0.5]{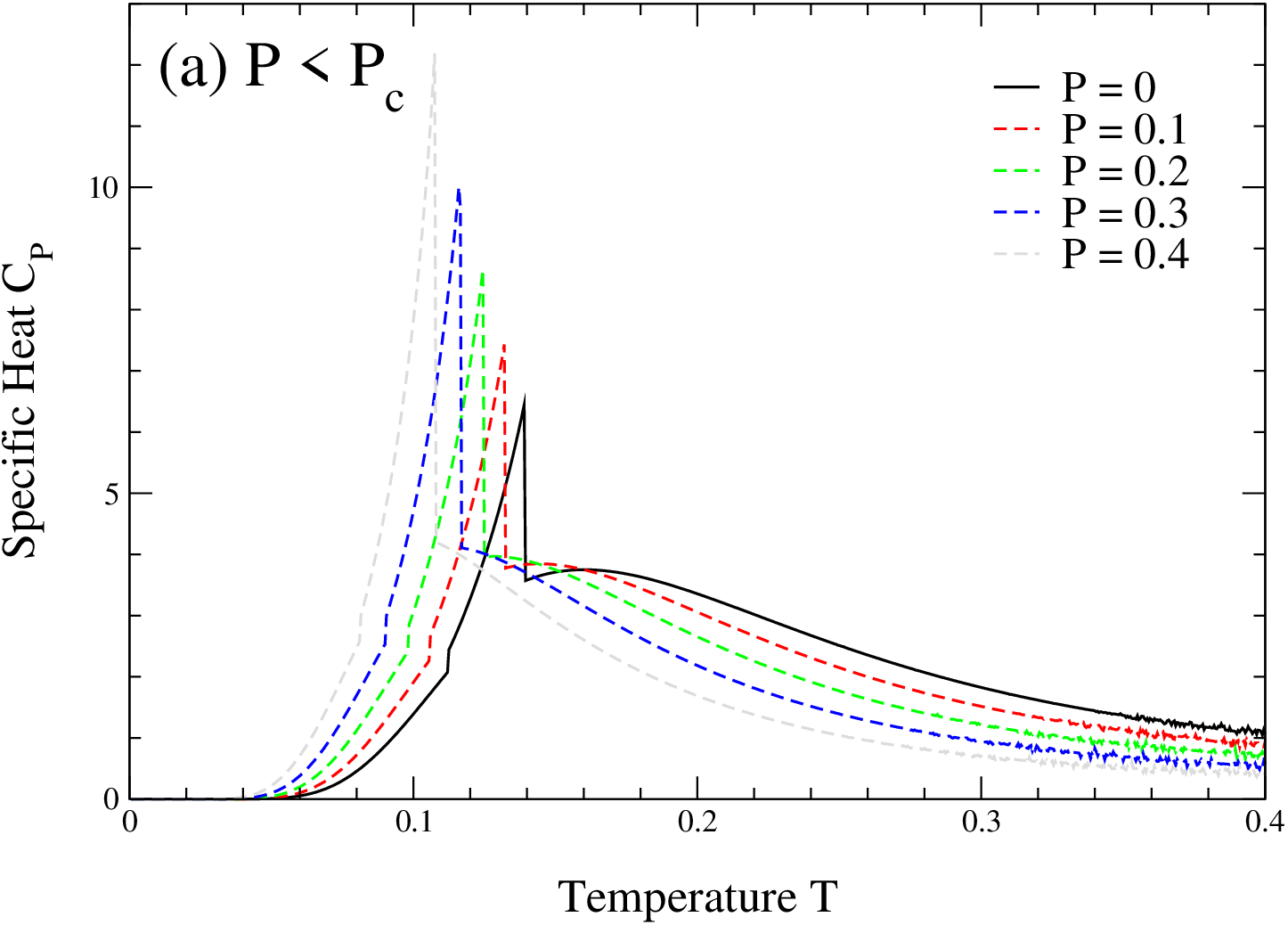}
\includegraphics[scale=0.5]{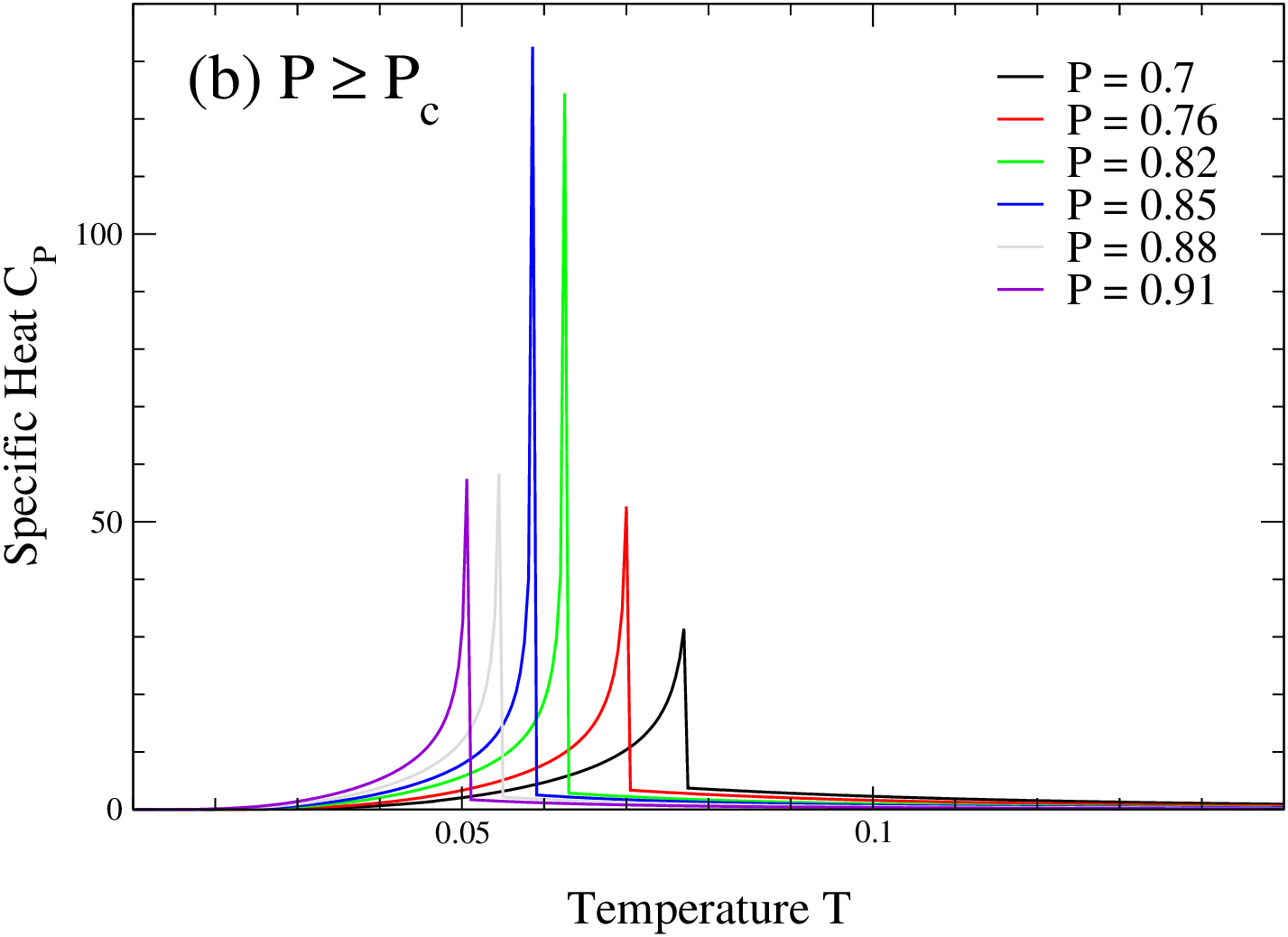}
\caption{{\it Mean Field Calculations:\/} Same as in Fig.~\ref{MC-CP} but
from mean field calculations (a) at $P<P_c^{\rm MF}$ and (b) at
$P$ approaching or larger than $P_c^{\rm MF}$. The mean field critical pressure is $P_c^{\rm
  MF}=0.82 \pm 0.04$.~\label{MF-CP}}
\end{figure}

\begin{figure}
\includegraphics[scale=0.5]{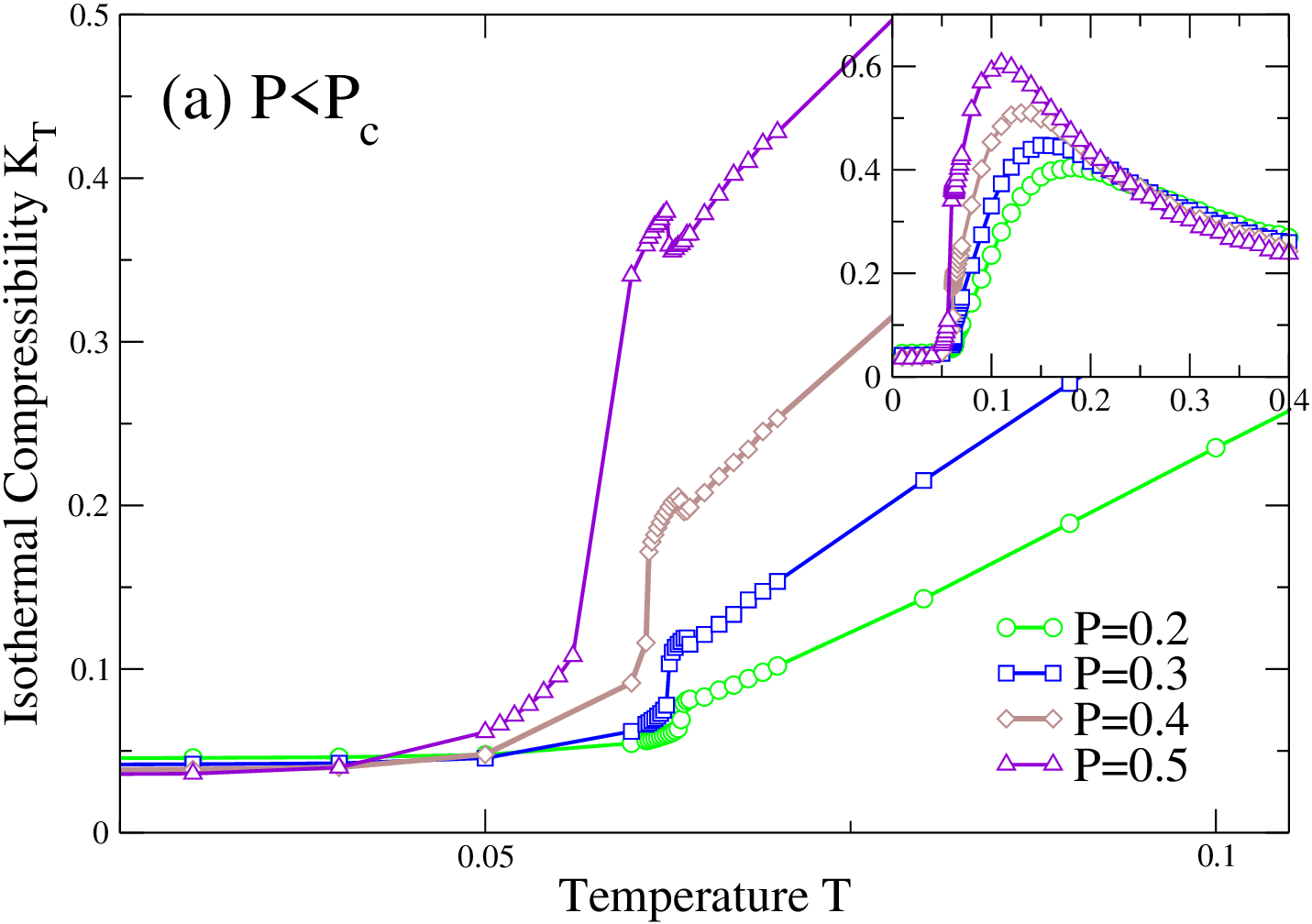}
\includegraphics[scale=0.5]{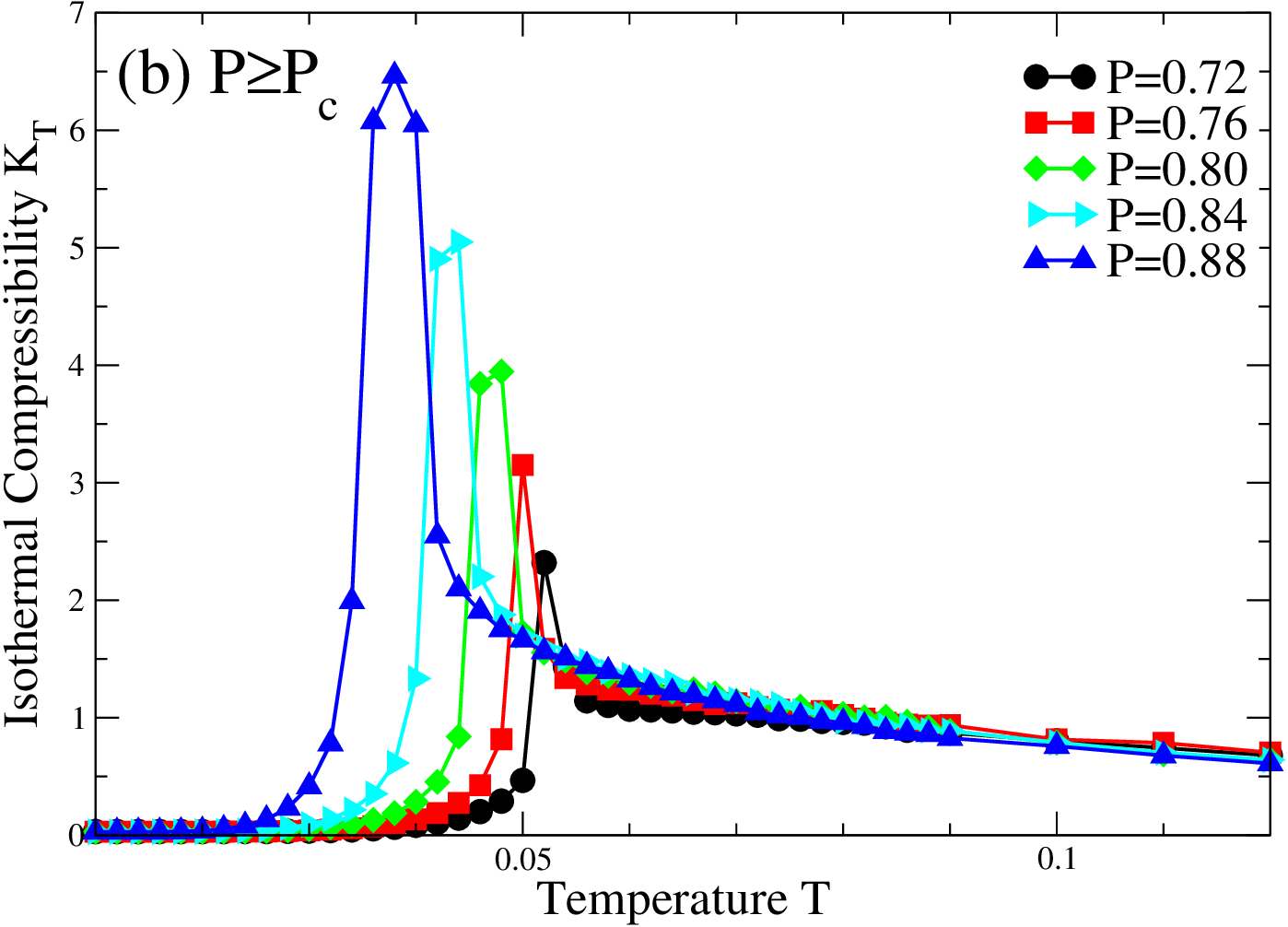}
\caption{{\it Monte Carlo Calculations:\/} (a) Temperature dependence of
  the isothermal compressibility $K_T$ along low pressure isobars with
  $P<P_c$. Inset: the high-$T$ broad maximum. (b) Same for $P \geq P_c$.
  \label{MC-KT}}
\end{figure}

At $P<P_c$, we find for $K_T$ a broad maximum that moves to lower $T$ upon increasing
$P$ (inset Fig.~\ref{MC-KT}a). This maximum occurs at higher $T$
with respect to the high-$T$ maximum that we find for $C_P$
(Fig.~\ref{MC-CP}a).  We also find a much smaller maximum of $K_T$ at a
$T$ that coincides, within the error bars, with the low-$T$ maximum of
$C_P$ (Fig.~\ref{MC-CP}a).

At $P>P_c$, we find a single maximum of $K_T$ (Fig.~\ref{MC-KT}b).
This maximum follows the LL phase transition and the low-$T$ maximum
of $C_P$ that we found for the same range of $P$ (Fig.~\ref{MC-CP}b).
The higher-$T$ maximum in $C_P$ for $P>P_c$ is not reflected in $K_T$, as
the model includes no volume change for the cooperative rearrangement
of the H bonds.

\begin{figure}
\includegraphics[scale=0.5]{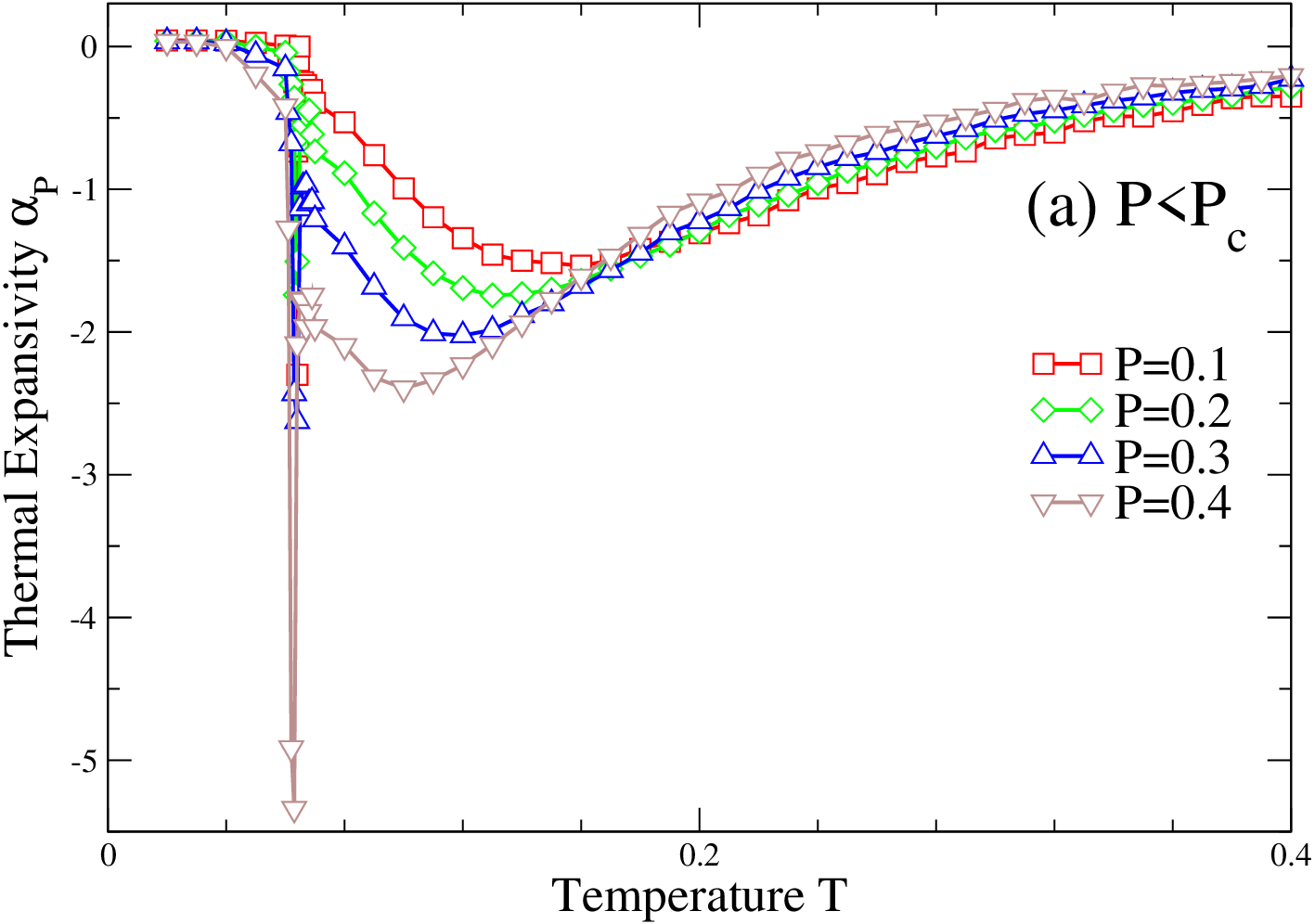}
\includegraphics[scale=0.5]{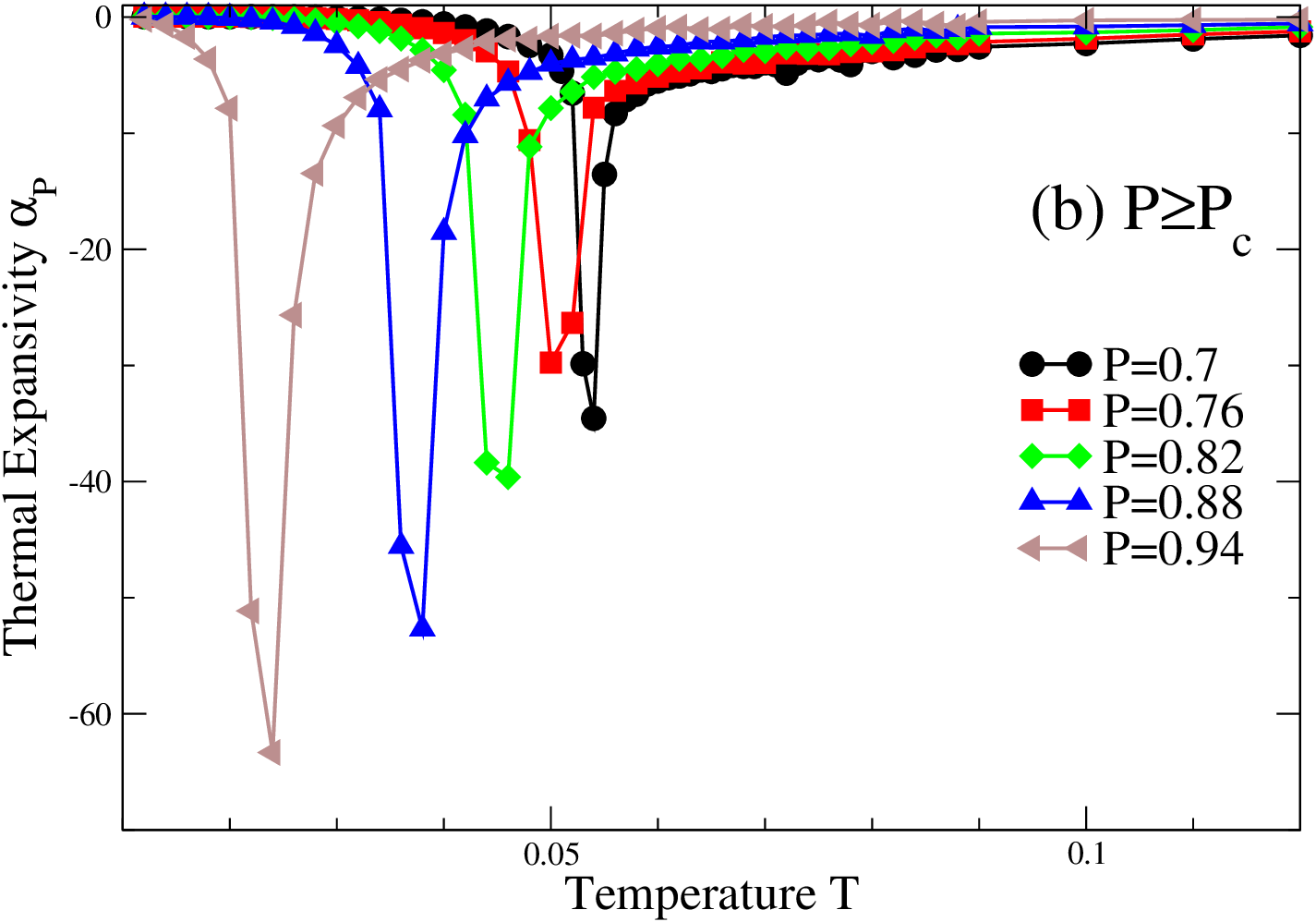}
\caption{{\it Monte Carlo Calculations:\/} (a) Temperature dependence of
  the thermal expansivity $\alpha_P$ along low pressure isobars with
  $P<P_c$. (b) Same for $P \geq P_c$.
  \label{MC-AP}}
\end{figure}

Our calculations of $|\alpha_P|$ at $P<P_c$ (Fig.~\ref{MC-AP}a) show a
behavior qualitatively similar to $C_P$ (Fig.~\ref{MC-CP}a), with a
broad maximum at higher $T$ and a sharp maximum at lower $T$.  The $T$
of the sharp maximum remains constant for increasing $P$, while the
higher-$T$ maximum decreases in $T$ with increasing $P$.  At $P>P_c$,
$|\alpha_P|$ shows a single maximum that follows the LL phase transition
(Fig.~\ref{MC-AP}b).  As with $K_T$, there is no equivalent to the
higher-$T$ maximum in $C_P$ at $P>P_c$, as the model includes no volume
change for cooperative rearrangement of the H bonds.

\subsubsection{Mean Field Calculations}

\begin{figure}
\includegraphics[scale=0.5]{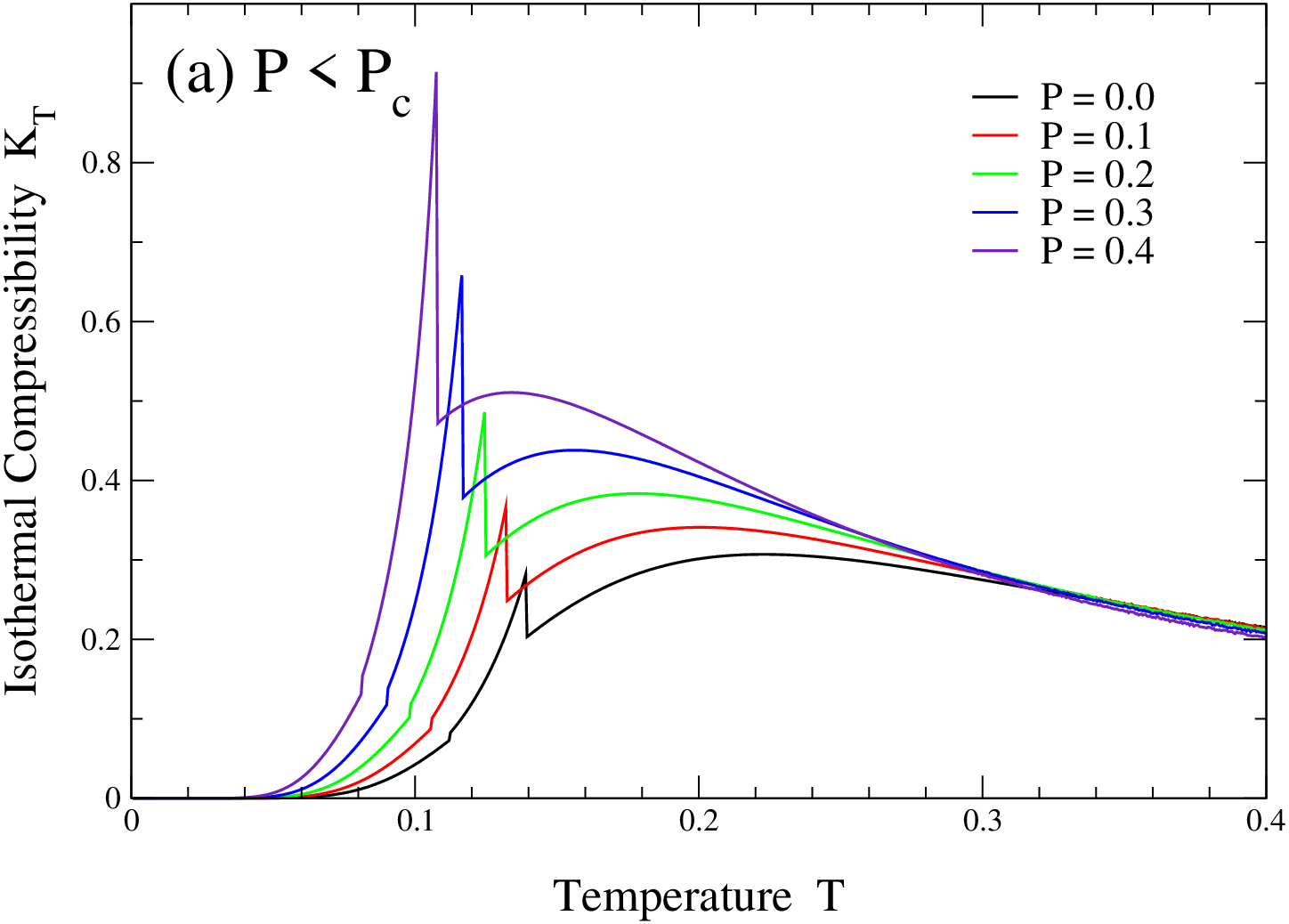}
\includegraphics[scale=0.5]{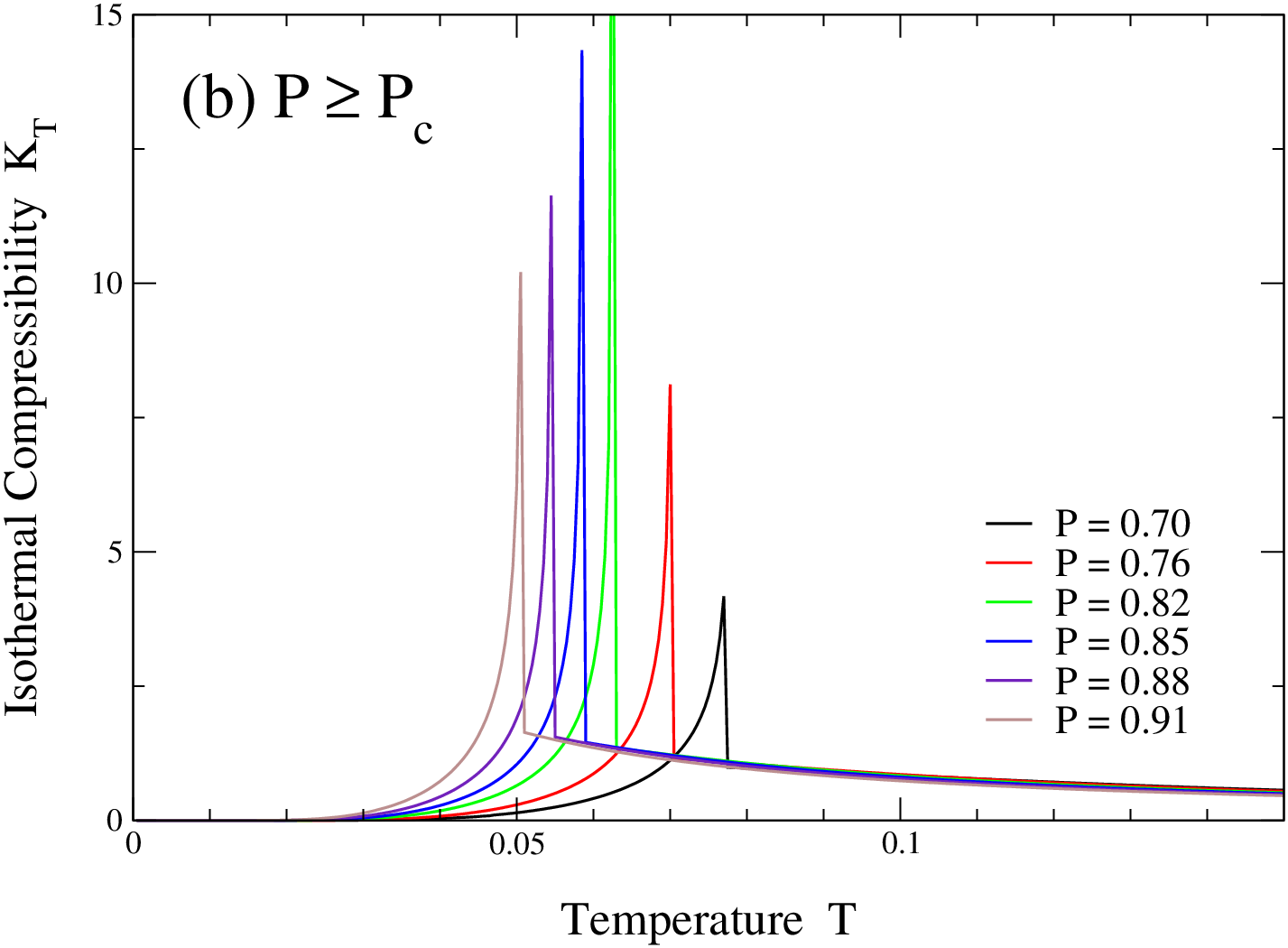}
\caption{{\it Mean Field Calculations:\/} (a) Temperature dependence of
  the isothermal compressibility $K_T$ along low pressure isobars with
  $P<P_c$. (b) Same for $P \geq P_c$.
  \label{MF-KT}}
\end{figure}

\begin{figure}
\includegraphics[scale=0.5]{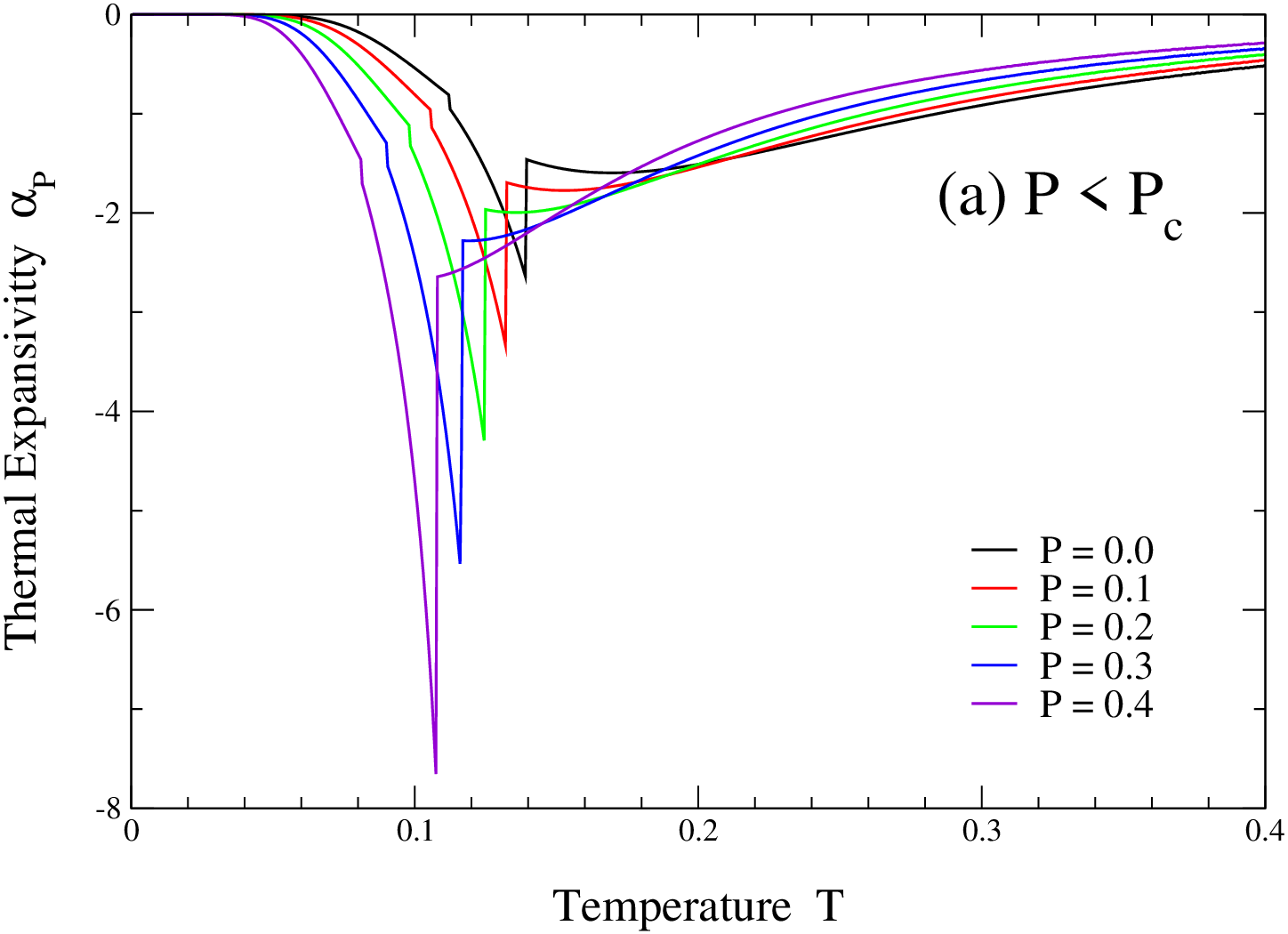}
\includegraphics[scale=0.5]{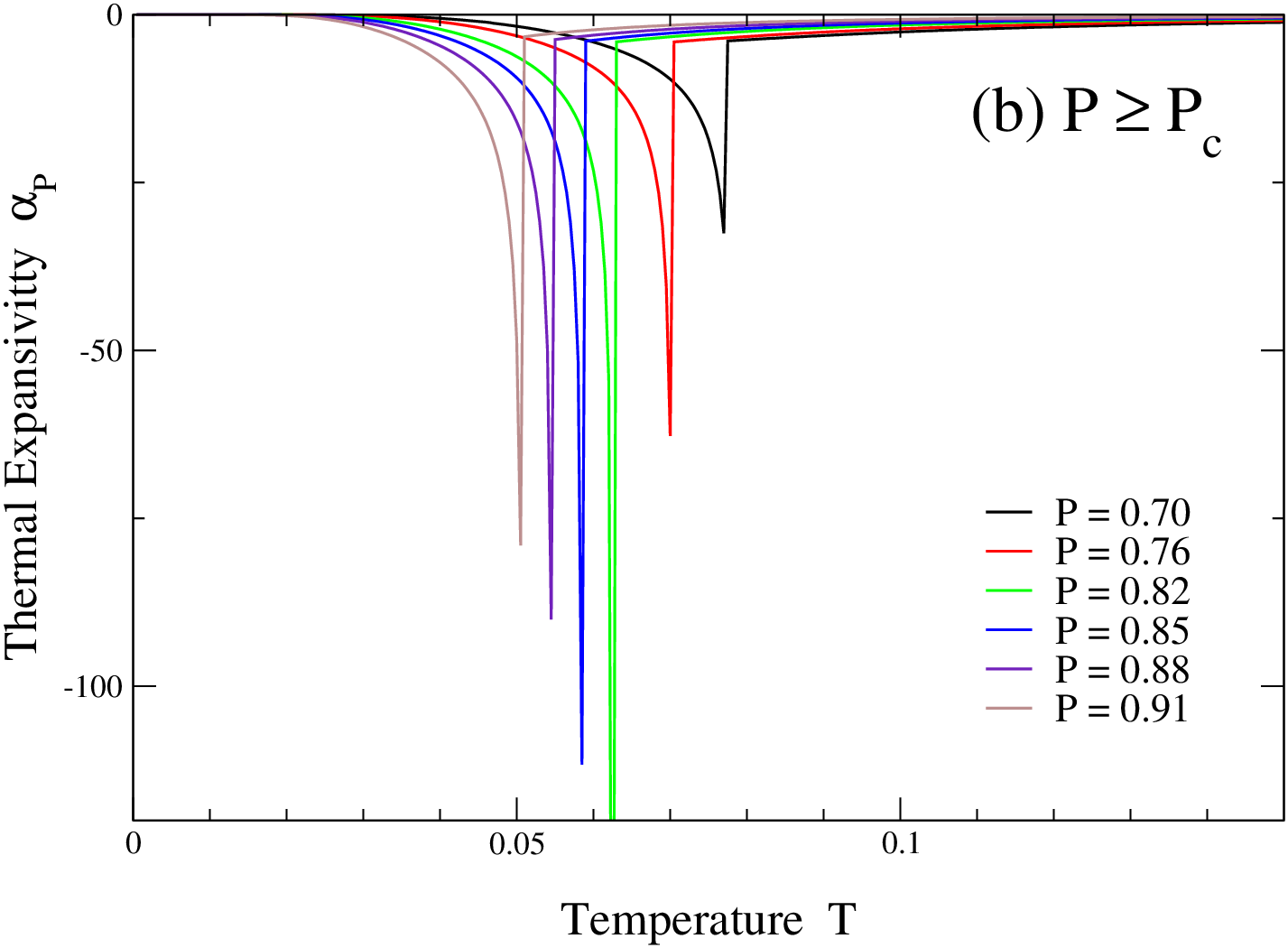}
\caption{{\it Mean Field Calculations:\/} (a) Temperature dependence of
  the thermal expansivity $\alpha_P$ along low pressure isobars with
  $P<P_c$. (b) Same for $P \geq P_c$.
  \label{MF-AP}}
\end{figure}

We can also calculate $K_T$ and $\alpha_P$ in the MF case.  At $P<P_c$,
$K_T$ (Fig.~\ref{MF-KT}a) and $|\alpha_P|$ (Fig.~\ref{MF-AP}a) exhibit
two maxima, which also move closer in $T$ with increasing $P$.  Because
these quantities are proportional to derivatives of the MF calculation
of the average number $\langle N_{\rm HB}\rangle$ of H bonds, as shown
in Eqs.~(\ref{MF-K})--(\ref{MF-a}), we interpret the maxima as in the case of
$C_P$, associating the high-$T$ maxima to the non-cooperative behavior
and the low-$T$ maxima to the cooperative interaction.

At $P>P_c$, $K_T$ (Fig.~\ref{MF-KT}b) and $|\alpha_P|$
(Fig.~\ref{MF-AP}b) show each a single maximum that follows the LL phase
transition. Also in this case the behavior is the same as what we find
for the MF calculations of $C_P$.

\section{Discussion}
\label{disc}

The $P$--$T$ phase diagram of the model displays the liquid-gas critical
point at the end of the first-order liquid-gas phase transition, the TMD
line, the LLCP at the end of the first-order LL phase transition, the
loci of maxima of the response functions that converge to each other,
approximating the Widom line
(Fig.~\ref{PHASE}a).  

\subsection{Low-pressure region, Widom line and glassy temperature}
Our results at low-$T$ show that the locus of the
two maxima of $C_P$, which we denote as $C_P^{1HB}$ and $C_P^{\rm
  coop}$, correlate well with the locus of maxima of $({\rm d}N_{\rm
  HB}/{\rm d}T)_P$ and $({\rm d}N_{\rm coop}/{\rm d}T)_P$, respectively
(Fig.~\ref{PHASE}b).  We find the same close correlation with the
structural changes of the H bond network for the loci of maxima of $K_T$
and $|\alpha_P|$ (Fig.~\ref{PHASE}b).

In a recent publication \cite{Mazza-PNAS11}, the proton relaxation time
$\tau$ for a monolayer of water adsorbed onto the surface of the protein
lysozyme was measured down to 150~K.  This relaxation time is caused by
charge defects moving along the H bond network, and thus probes
the time-scale of H bond reorientation.  It was found that the
$T$-dependence of $\tau$ exhibits two crossovers in the region of the
phase diagram in which two $C_P$ maxima are found in the present cell
model.

\begin{figure}
\includegraphics[scale=0.5]{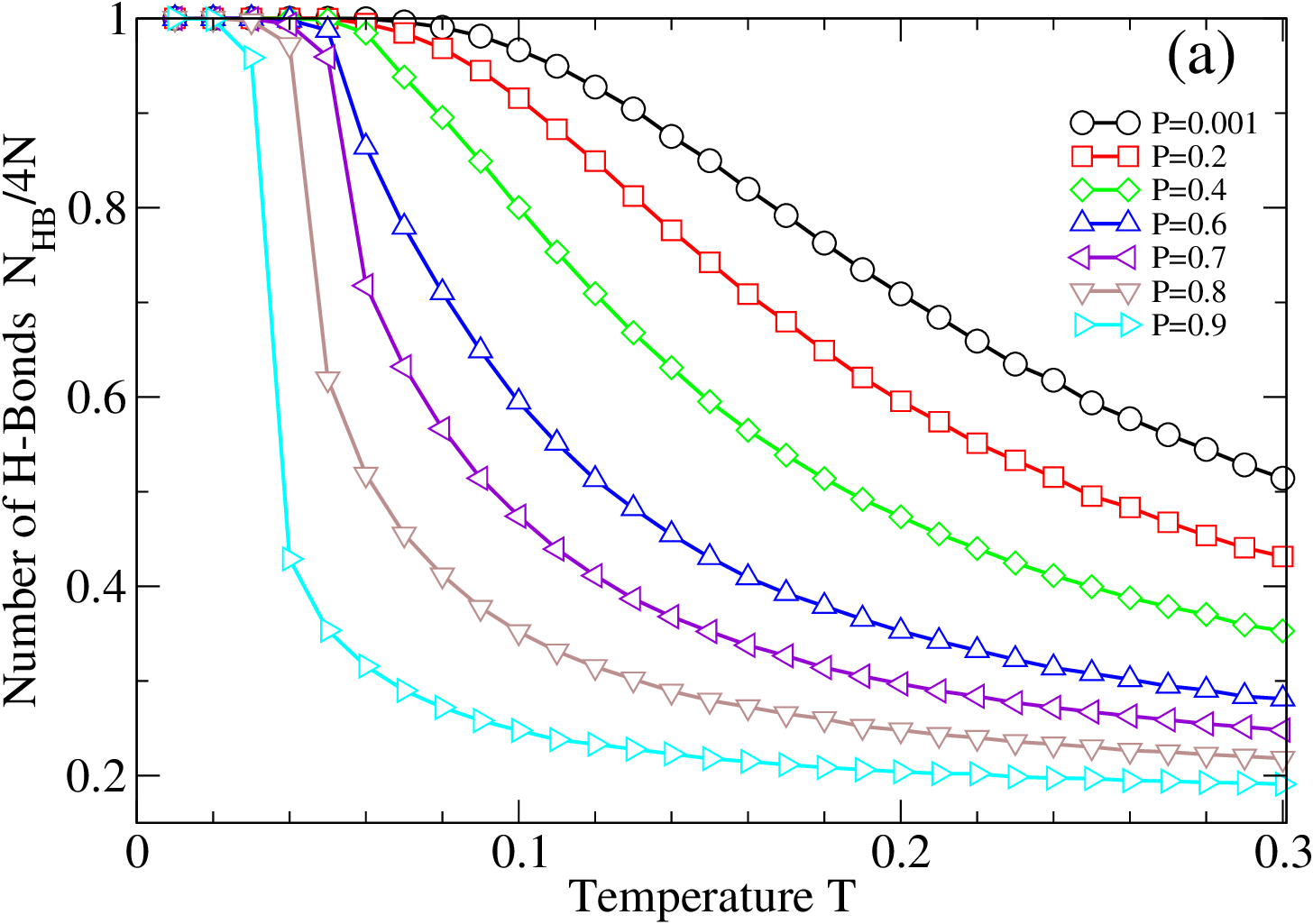}
\includegraphics[scale=0.5]{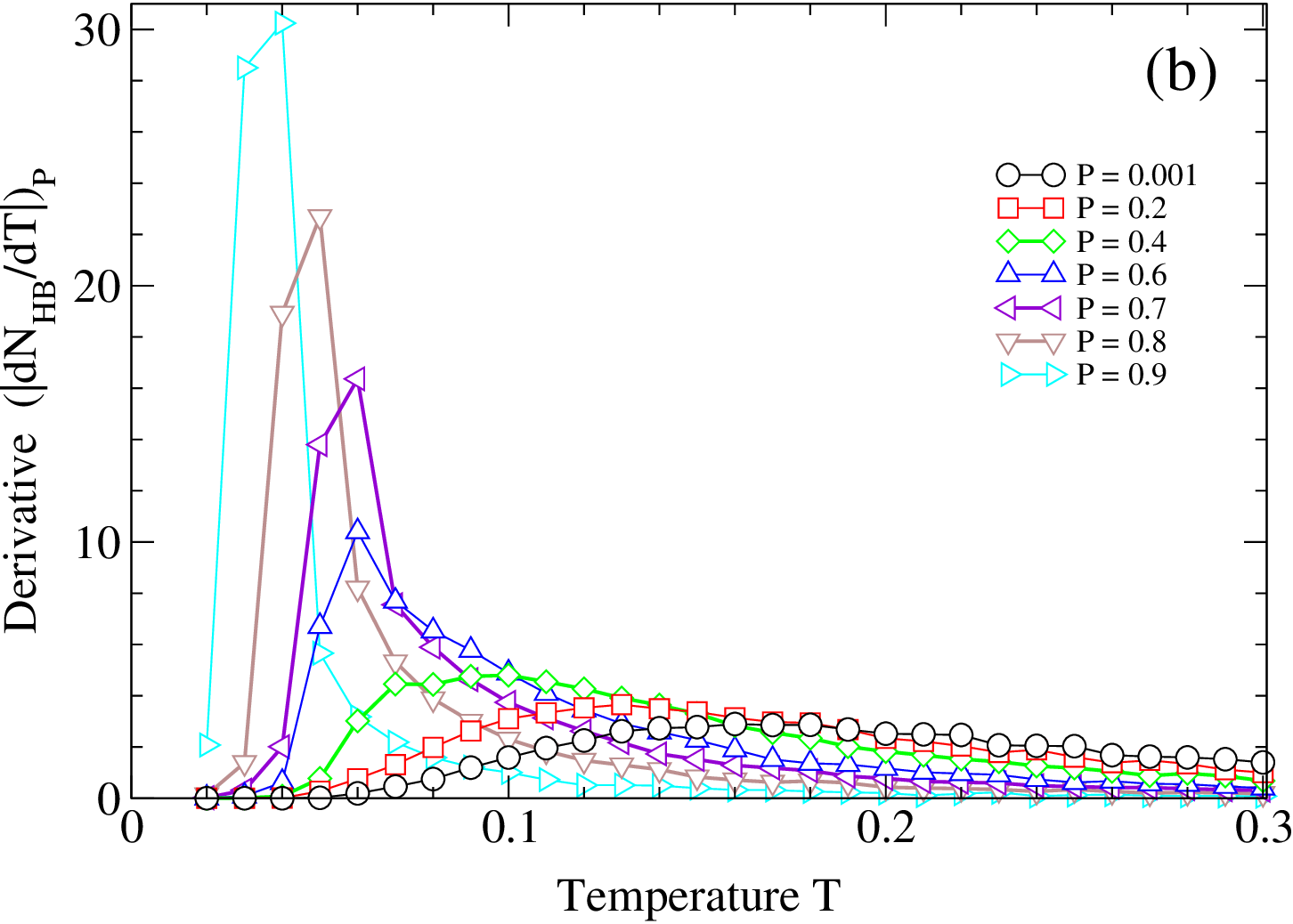}
\caption{{\it Monte Carlo Calculations:\/} Temperature dependence of (a)
  the number of H bonds $N_{\rm HB}$, divided by the total number of
  possible H bonds $4N$, and (b) its $T$-derivative $(\partial \langle
  N_{\rm HB} \rangle /\partial T)_P$ as a function of $T$ for different
  isobars. The temperatures of the maxima of the derivative overlap with
  the temperatures of the maxima of $C_P^{\rm 1HB}$ in Fig.~\ref{MC-CP}.
\label{MC-NHB}}
\end{figure}

The physical interpretation of these thermodynamic and dynamic results
is straightforward. By decreasing $T$ at low $P$ the water molecules
form an increasingly large number of H bonds. The largest structural
change associated with this formation of H bonds is marked by the maximum
in $(\partial \langle N_{\rm HB} \rangle /\partial T)_P$
(Fig.~\ref{MC-NHB}).  Note that at any $P$ this maximum occurs where the
probability $p_B\equiv N_{\rm HB}/(4N)$ of forming a H bond is
approximately $p_B=0.8$ \cite{KumarJPCM2008} (Fig.~\ref{MC-NHB}).
Although under these conditions the number of H bonds is macroscopic,
they form independently, often with different relative orientations
(bonding states). Thus at this stage the H bond regions that are
thermodynamically correlated have a characteristic but finite size
\cite{BIF2012,FBS2011}.  Nevertheless, the formation of these finite
clusters of correlated H bonds implies a change in their dynamics, with
a crossover from a high-$T$ non-Arrhenius dynamics to a new regime at
$T$ below the maximum of $(\partial \langle N_{\rm HB} \rangle /\partial
T)_P$ and response functions, whose characteristics we discuss in the following.

As observed in Ref.~\cite{KumarPRL2008}, at any $P$ the crossover occurs
when the H bond relaxation time reaches a characteristic value. This
value has been estimated to be $\tau=10^{-4}$~s in
Ref.~\cite{Mazza-PNAS11}, based on a comparison with dielectric
spectroscopy experimental results for water adsorbed on lysozyme powder
at a low hydration level (0.3 g H$_2$O/g dry protein). This time scale
is seven orders of magnitude larger than the characteristic single molecule
rotational relaxation time $\tau_{\rm rot}\simeq 28$~ps
\cite{kumarPRE2006} at the same temperature of the crossover, $T\approx
252$~K. Thus the crossover is associated with a relaxation mode that
involves more than one molecule, consistent with our interpretation
based on finite clusters of correlated H bonds with a characteristic
average size.

Although the first analysis suggested that the dynamics below the broad
maximum of $C_P$ is Arrhenius \cite{KumarPRL2008,KumarJPCM2008}, further
investigations extended to lower $T$ have shown that it is non-Arrhenius
\cite{Mazza-PNAS11}.  The dynamics becomes Arrhenius only at $T$ below
the lower-$T$ maximum of $C_P$ \cite{Mazza-PNAS11}.

\begin{figure}
\includegraphics[scale=0.5]{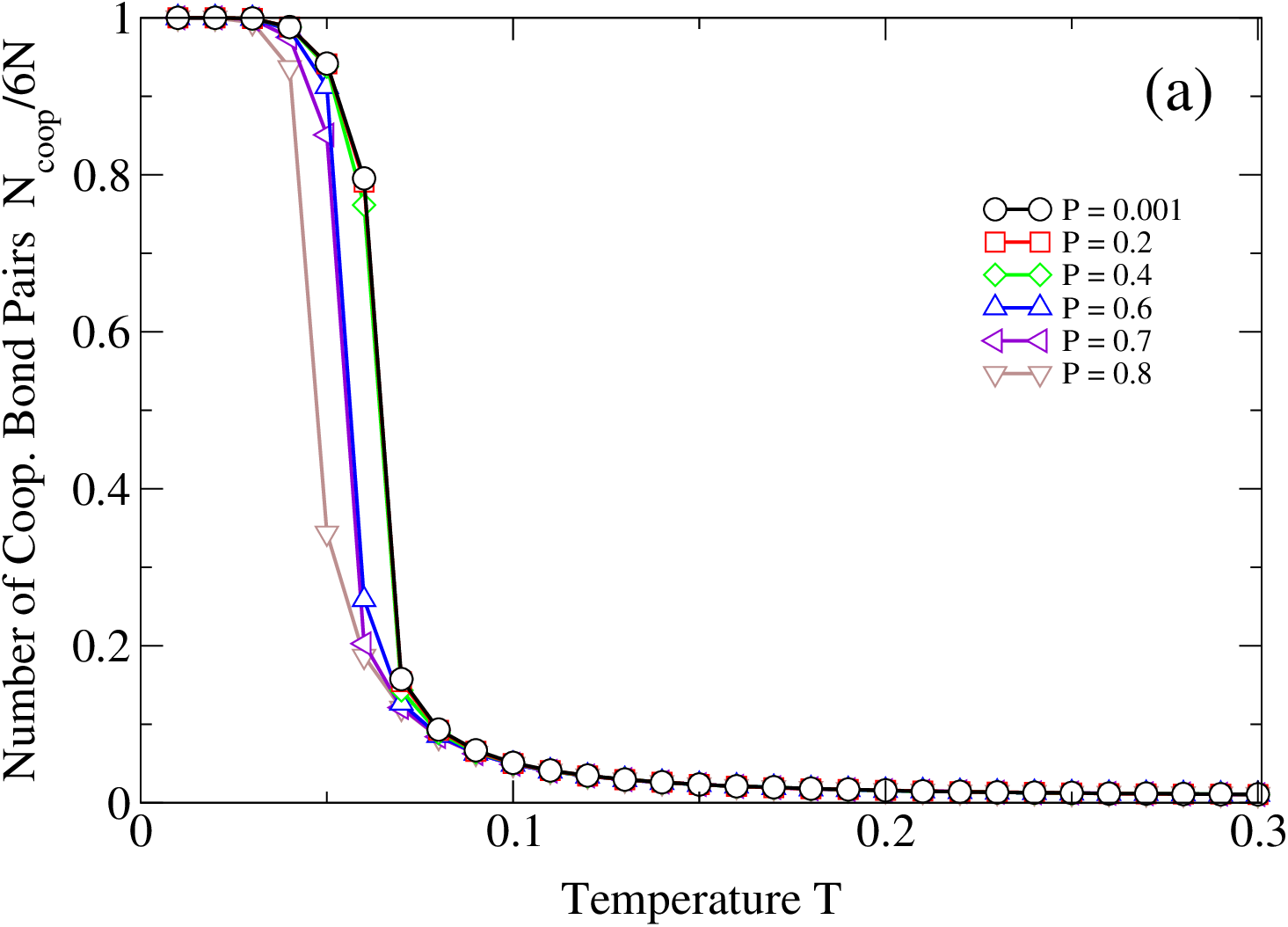}
\includegraphics[scale=0.5]{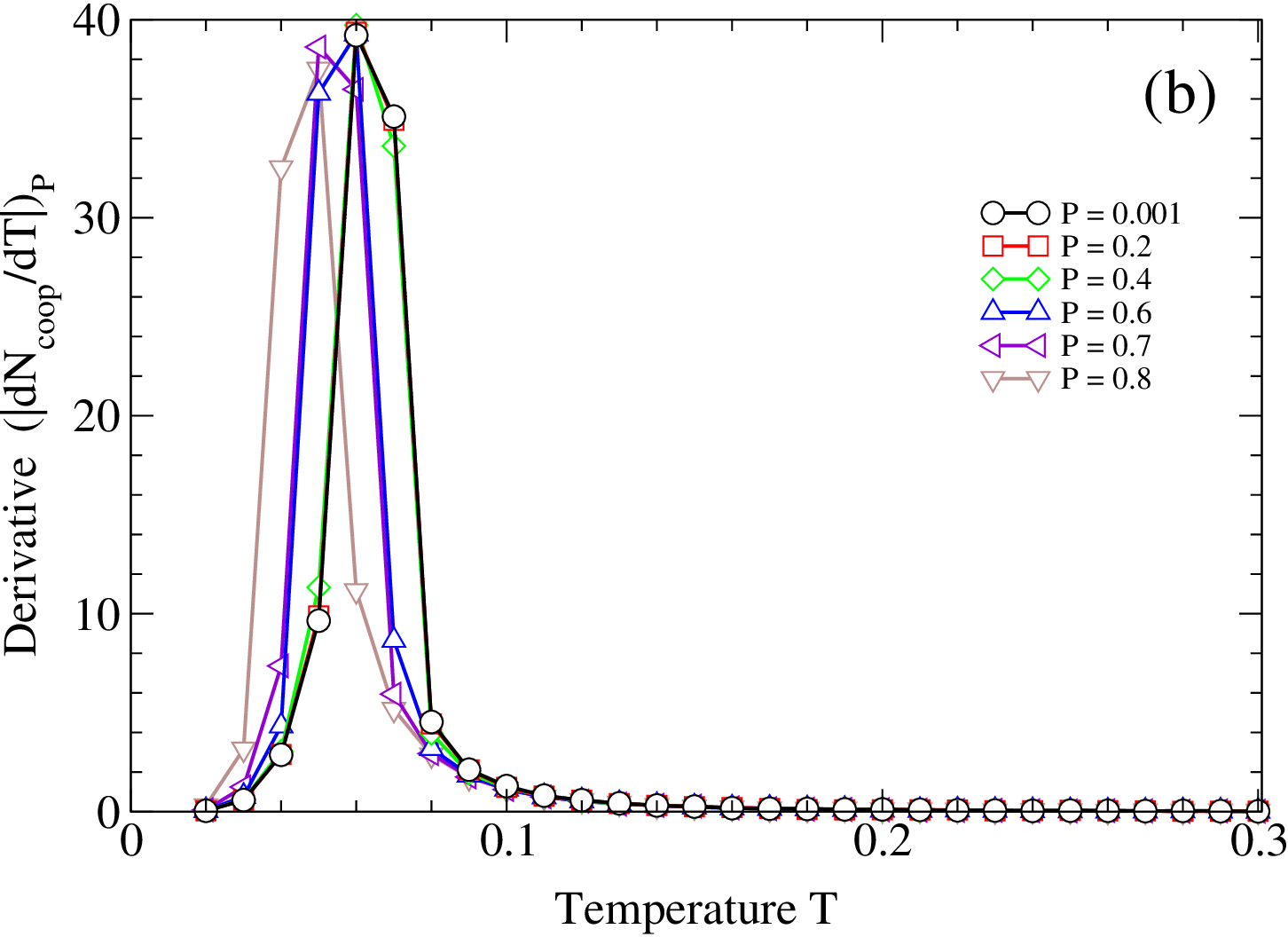}
\caption{{\it Monte Carlo Calculations:\/} Temperature dependence of (a)
  the number $N_{\rm coop}$ of cooperative pairs of H bonds formed by
  the same molecule, divided by the total number of possible H bonds
  pairs on the same molecule $6N$, and (b) its $T$-derivative $(\partial
  \langle N_{\rm coop} \rangle /\partial T)_P$ as a function of $T$ for
  different isobars. The locus of the maxima of this derivative
  overlaps with the locus of the maxima of $C_P^{\rm coop}$ in
  Fig.~\ref{MC-CP}.
\label{MC-NIM}}
\end{figure}

As discussed in Ref.~\cite{Mazza-PNAS11}, we understand that the dynamic
behavior is not Arrhenius between the two maxima of $C_P$ because at
these temperatures the system has an activation energy that is
$T$-dependent. This happens because interfaces with high free energy
costs appear among the different clusters of correlated H bonds.  By
decreasing $T$ at constant $P$, the water many-body interaction,
parametrized with $J_\sigma$ in the model, induces a cooperative
rearrangement of the H bonds, gradually eliminating the interfaces among
the clusters. This restructuring of the H bond clusters reaches its
maximum when the $(\partial \langle N_{\rm coop} \rangle /\partial T)_P$
reaches its maximum (Fig.~\ref{MC-NIM}), resulting in the second maximum of
$C_P$ (Fig.~\ref{MC-CP}).

At this stage the cluster size of the correlated H bonds increases.  As
a consequence, a majority of the water molecules now have four H bonds
that satisfy the cooperative interaction. Hence the dynamic behavior is
dominated by processes with a characteristic activation energy. It has
been shown by Mazza et al. that this activation energy is consistent
with the average energy necessary to break a H bond in a cooperatively
ordered environment \cite{Mazza-PNAS11}. Thus the dynamics has a second
crossover below the low-$T$ maximum of $C_P$, this time to an Arrhenius
behavior.

As consequence of the increase of the cluster size of the correlated
H bonds, we identify, at low $P$ ($P<P_c$), the temperature of the
low-$T$ crossover with the Widom line associated with the LL phase
transition \cite{Limei-PNAS05,Franzese-JPCM07}.
This suggests that previous works identifying a LL Widom line in
supercooled water
\cite{Limei-PNAS05,Franzese-JPCM07,JPCM08,Kumar2007,Mallamace2008,Abascal2010} 
may have only observed the higher--$T$ line of response maxima, the
true Widom line residing at lower temperature. 


The fact that this maximum occurs at an approximately constant value of
$p_{\rm coop}\equiv N_{\rm coop}/(6N)\simeq 0.5$ at any $P$
(Fig.~\ref{MC-NIM}) implies that the cluster size of the correlated
H bonds is independent of $P$. Analogous to what we have found for the
first maximum of $C_P$, we expect this equal-cluster-size property at
the crossover to correspond to an equal-characteristic-time at the
crossover. Although at the present time we can only speculate this
isocronic (equal time) property at the low-$T$ crossover, based on our
comparison with the dielectric experimental results we predict the
characteristic time to be of the order of 2~s \cite{Mazza-PNAS11}. This
estimate is consistent with the idea that this crossover, and the Widom
line, occurs at temperatures $T_W(P)$ that are above, but not far from,
the glassy temperature $T_g(P)$ of the H bonds, defined as the
temperature at which the H bond characteristic relaxation time exceeds
100~s. This conclusion is consistent with what has been recently found
in long-time simulations for ST2 bulk water \cite{Kesselring2012}.

By increasing $P$, but with $P<P_c$, the increase in volume due to
H bond formation contributes significantly to the enthalpy, making the
H bonds unfavorable. Thus H bonds form at a $T$ that decreases with
increasing $P$, and the maximum of $({\rm d}N_{\rm HB}/{\rm d}T)_P$
shifts to a lower $T$.  On the other hand, the $T$ at which $({\rm
  d}N_{\rm coop}/{\rm d}T)_P$ is at a maximum remains approximately
constant with $P$, as there is no volume cost associated with the
cooperative rearrangement.  Therefore the two maxima in $C_P$ move
closer in $T$ when $P<P_c$ increases (Fig.~\ref{PHASE}).

The fact that the maximum of $({\rm d}N_{\rm coop}/{\rm d}T)_P$ does not
depend much on $P$ implies that the cooperative maximum of $C_P$ and the
Widom line of the LL phase transition also do not depend strongly on
$P$. This prediction that the Widom line behavior is a weak function of $P$
has been confirmed recently by a fitting of the thermodynamic data based
on the LLCP hypothesis \cite{Holten}.

Although $K_T$ and $\alpha_P$ do not explicitly depends on $N_{\rm
  coop}$, they also show as $C_P$ a maximum at low $P$ and low $T$
that follows the maximum of $({\rm d}N_{\rm coop}/{\rm d}T)_P$. We
understand this behavior as a consequence of the fact that $N_{\rm
  HB}$, from which $K_T$ and $\alpha_P$ depend, is affected by a large
change of $N_{\rm  coop}$, because by increasing the number $N_{\rm
  coop}$ of bonding indices of the same molecules in the same state,
also the number $N_{\rm  HB}$ of H bonds between different molecules
increases at low $T$, spreading the local order at a distance that is
of the order of the correlation length $\xi$. 

Next, we discuss the difference of our findings with those
presented for an isotropic potential describing an anomalous liquid,
where two maxima of $C_P$ at different temperatures were observed in
the supercritical region with respect to the LLCP \cite{Sergey09}. In
this case the low-$T$ maximum is a consequence of
the out-of-equilibrium dynamics of the system and is related to the
glass transition temperature $T_g(P)$ of the liquid, as emphasized by
the cooling-rate dependence of the low-$T$ maxima of  the
isotropic potential $C_P$ \cite{Sergey09}. In our case, as discussed above,
all the maxima occur at $T>T_g(P)$. Furthermore, our calculations can
be equilibrated at any $T$ thanks to the efficient Wolff algorithm,
characterized by  short MC correlation times \cite{mazzaCPC},
consistent with our MF analysis that is, by definition, at equilibrium.

Finally, we observe that the molar heat capacities of the water
confined within nano-pores of silica MCM-41 measured with adiabatic
calorimetry for pores with 1.8 nm diameter shows two maxima
\cite{Oguni-08}. The 
low-$T$ maximum in this case has been associated to the 
crystallization of part of the water in the pore \cite{Oguni-08}. In
our case, we can exclude any crystallization effect by direct analysis
of the dynamics of the system, consistent with previous simulations
\cite{Zangi, Kumar}. A diffusion analysis has shown that our
system is subdiffusive at the Widom line temperatures  
\cite{JPCB2011,PRER2012}, but not crystalline.

\subsection{The liquid-liquid critical point and the high-pressure region}

\begin{figure}
\includegraphics[scale=0.4]{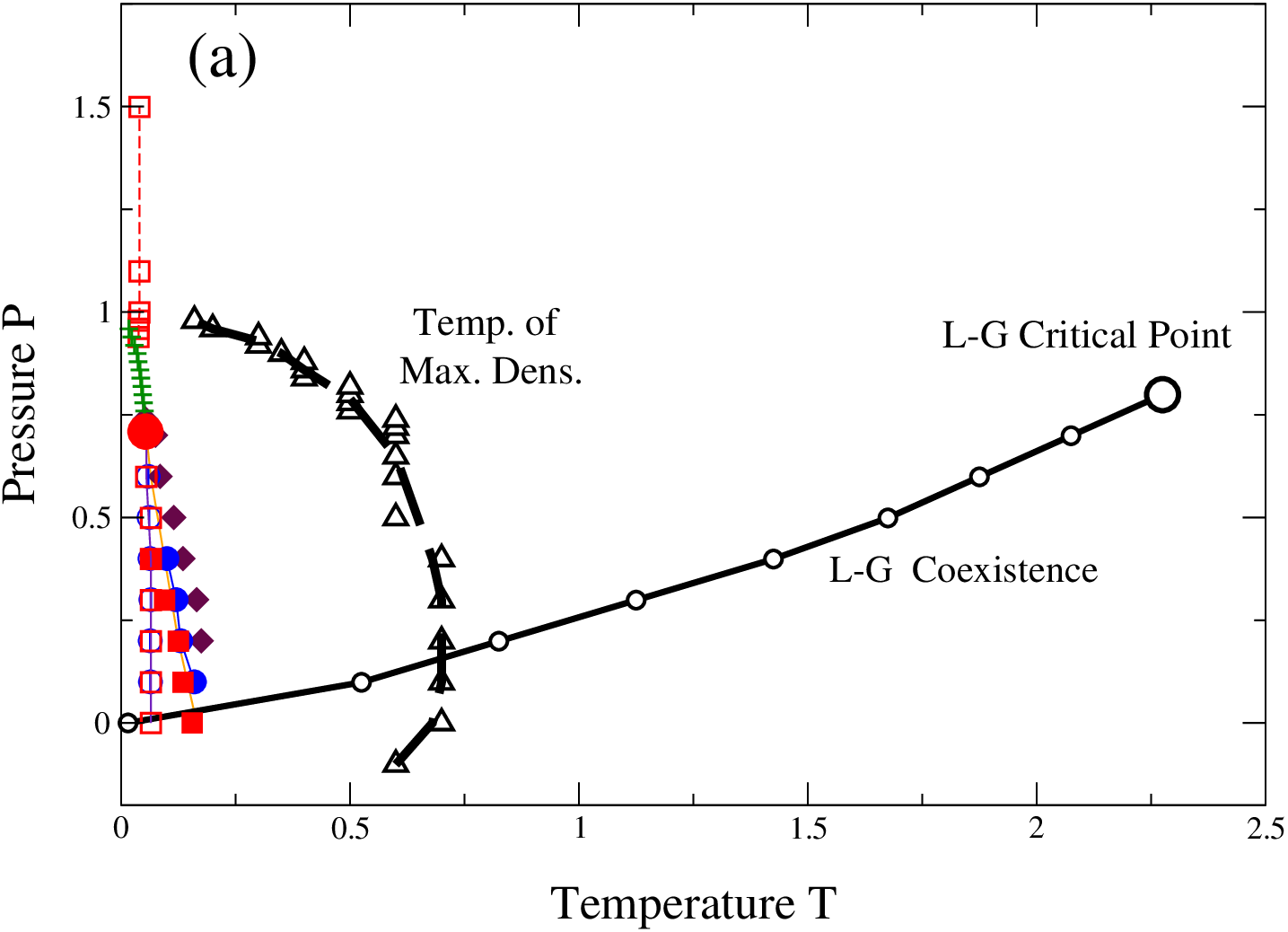}
\includegraphics[scale=0.4]{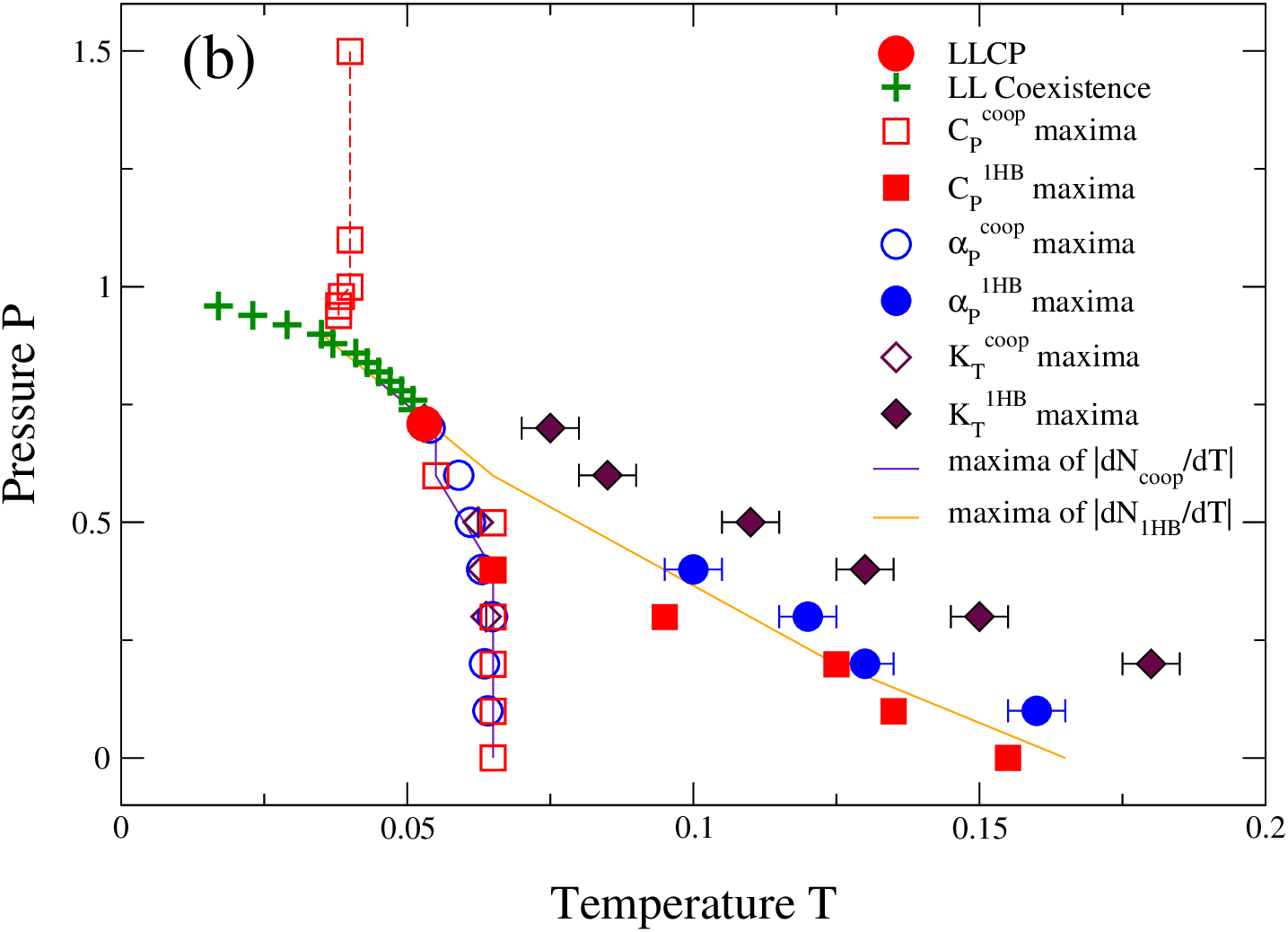}
\caption{{\it Monte Carlo Calculations:\/} (a) Phase diagram showing the
  liquid--gas coexistence (continuous thick with circles) line, the
  temperature of maximum density (TMD, dashed thick with triangles)
  line, the LL coexistence line (pluses) ending in the  LLCP
  (large full circle), and other lines
  that are better described in the second panel.  (b) Magnification of
  the phase diagram at low $T$.
  At $P$ well below the LLCP, $C_P^{\rm coop}$ (empty 
  squares), $\alpha_P^{\rm coop}$ (empty circles) and $K_T^{\rm coop}$
  (empty diamonds), on one hand, and $C_P^{\rm 1HB}$ (filled squares),
  $\alpha_P^{\rm 1HB}$ (filled circles), and $K_T^{\rm coop}$ (filled 
  diamonds), on the other hand, are maximal at different $T$.
  All the loci of maxima of $C_P$, $K_T$, and
  $|\alpha_P|$ converge toward the LLCP, together with the loci of
  maximal $({\rm d}N_{\rm coop}/{\rm d} T)_P$ (dark thin line) and
  $({\rm d} N_{\rm 1HB}/{\rm d}T)_P$ (light thin line), delimiting a
  region in the vicinity of the LLCP that approximates the Widom line.  For $P>P_c$,
  all the loci coincide within the error with the LL coexistence line,
  but the locus of maximum $C_P^{\rm coop}$. If not shown, error bars
  are smaller that the symbol size.}
\label{PHASE}
\end{figure}

When $P<P_c$, the response functions $C_P$, $K_T$, and $\alpha_P$
exhibit maxima that occur at temperatures that approach each other as
$P$ increases (Fig.~\ref{PHASE}). For the same state points, the absolute values
of their maxima increase and reach their maximum value when the loci of
their maxima in the plane $P$-$T$ merge (Fig.~\ref{PHASE}b). This is the
behavior we would expect at the critical point in a finite system, where
by definition the value of the response functions cannot diverge. We
therefore locate the LLCP ($P_c$, $T_c$) at the thermodynamic
point where the maxima of the response functions and of 
$({\rm d}N_{\rm HB}/{\rm d}T)_P$, and $({\rm d}N_{\rm
  coop}/{\rm d}T)_P$ merge (Fig.~\ref{PHASE}b).

The LLCP occurs where both the structural change associated with $N_{\rm
  HB}$ and that associated with $N_{\rm coop}$ have their maxima. This
is because at this state point the macroscopic increase in the number of
H bonds is amplified by their simultaneous rearrangement into a
cooperative configuration. This  gives rise to a cooperative
phenomenon that encompasses the entire system and causes a continuous
change from a high-$T$, highly disordered, high-energy, high-density
liquid (HDL) phase with a few H bonds, to a low-$T$, more ordered,
low-energy, low-density (LDL) liquid phase with many cooperatively
ordered H bonds.

When $P>P_c$, H bonds form at a $T$ that continues to decrease with
increasing $P$, due to their enthalpic cost for the density decrease.
This cost can be counterbalanced only by a large energy gain and entropy
loss. This condition is realized only when the H bonds adopt the
orientation that minimizes the many-body, cooperative interaction. Thus
when a macroscopic number of H bonds is formed, the free energy of the
system experiences a discontinuous change that results in a first-order
phase transition. This transition is marked by sharp maxima in $C_P$,
due to the $C_P^{1HB}$ component, $K_T$, and the absolute value of
$\alpha_P$ (Fig.~\ref{PHASE}).

At $T$ higher than the LL phase transition, the few H bonds that have
formed will themselves rearrange locally, resulting in a maximum of
$({\rm d}N_{\rm coop}/{\rm d}T)_P$.  This in turn implies a maximum in
$C_P$ due to the $C_P^{\rm coop}$ component (Fig.~\ref{PHASE}).

\section{Summary and Conclusions} 

A microscopic cell model for a confined water monolayer that exhibits a
LL phase transition and LLCP shows two maxima in $C_P$ at very low $T$.
We decompose $C_P$ as a sum of two terms, $C_P^{\rm 1HB}$ and $C_P^{\rm
  coop}$, each of which is responsible for one of the maxima in $C_P$.
We find that $C_P^{\rm 1HB}$ is caused by fluctuations in the formation
of H bonds, while $C_P^{\rm coop}$ is caused by fluctuations arising
from the cooperative interaction among H bonds. We find two maxima
also in both $K_T$ and $\alpha_P$, occurring at the same temperatures
as the two maxima of $C_P$. 

A physical picture emerges from these results.  The liquid at ambient
conditions contains few molecules that form strong H bonds with
neighboring molecules.  At low $P$, upon cooling, the entropy cost of
bond formation contributes less to the free energy, and bonds form
independently.  The rate of H bond formation, $(\partial \langle N_{\rm
  HB} \rangle /\partial T)_P$, reaches a maximum that is reflected in
$C_P$ by the maximal contribution of $C_P^{\rm 1HB}$.  With further
cooling at low $P$, the H bonds reorganize into a more cooperative
arrangement.  This molecular reorganization is dominated by the small
energy scale of the many-body interactions among the water molecules,
occurring at the $T$ where $C_P$ has the maximal contribution of
$C_P^{\rm coop}$.

Recent experimental results on a water monolayer hydrated lysozyme
powder are consistent with the occurrence of the two $C_P$ maxima, as
discussed in Ref.~\cite{Mazza-PNAS11}. In the experiment, the H bond
dynamics show two crossovers at separate temperatures. These are
reproduced by the model, which shows that they are a consequence of the
two $C_P$ maxima and the two associated structural changes. Thus the
measurement of the two crossovers strongly supports the existence of
two $C_P$ maxima in water at low $P$.

At higher $P$ the volume contribution to the enthalpy becomes more
relevant. It affects the formation of the H bonds, causing the decrease
of the temperature of maximum $C_P^{\rm 1HB}$. On the other hand, the
volume change due to the cooperative rearrangement is assumed to be
negligible in our approach, and $P$ has no effect on the locus of
maximum $C_P^{\rm coop}$.  The consistency of our predictions with a
recent analysis of available experimental data \cite{Holten} supports
our assumption.

The $P$-dependence of the locus of maximum $C_P^{\rm 1HB}$, opposite to
the $P$-independence of the locus of maximum $C_P^{\rm coop}$, allows us
to predict that the two maxima of $C_P$ move closer in $T$ with
increasing $P$ for $P\ll P_c$, to separate in $T$ with increasing $P$
for $P\gg P_c$, and to cross each other in the vicinity of $P_c$. For
$K_T$ and $\alpha_P$, instead, the two maxima are present only for
$P\ll P_c$, merging approaching $P_c$.

These predictions enable us to discriminate among the scenarios that
have been proposed for the phase diagram of water, including the SF,
LLCP, and CPF (or its equivalent SL) scenarios.  In our study of the
phase diagram of an adsorbed monolayer of water we find that each
scenario predicts a unique behavior of $C_P$,  $K_T$ and $\alpha_P$ at
supercooled $T$. In 
particular, a measurement of the two crossovers in the dynamics
\cite{Mazza-PNAS11}, interpreted as a consequence of the two maxima of
$C_P$, rules out the SF scenario, because in the SF scenario there is only one
maximum and, as a consequence, only one crossover.

Our predictions of different pressures also suggest an experimental test
to discriminate between the LLCP and the CPF scenarios, i.e., a
measurement of the $C_P$, or $K_T$ or $\alpha_P$ maxima at several pressures
around ambient pressure 
under supercooled conditions. Indeed, at low $P$ the two maxima in the
response functions
should approach each other if the LLCP scenario holds.
If, instead, the CPF scenario is verified or if the LLCP occurs at
a pressure below those investigated, the two maxima of $C_P$ should
separate further, while $K_T$ and $\alpha_P$ should have only one maximum.
 Thus by measuring how these maxima move in $T$
at several $P$ we could determine whether (i) the two maxima of $C_P$, or
$K_T$ or $\alpha_P$  merge at
positive $P$, giving a lower-bound estimate of the LLCP, (ii) the two
maxima of $C_P$ merge at
negative $P$ above the limit of stability of the liquid with respect to
the gas, giving an upper-bound estimate of the LLCP, or (iii) the two
maxima of $C_P$ do not
merge before the liquid-to-gas limit of stability, ruling out the LLCP
and supporting the CPF scenario. 

\section{Acknowledgments}

We thank F.~Mallamace, S.~Sastry and E.~Strekalova for helpful
discussions. We acknowledge the support of NSF grants  CMMI-1125290,
CHE-0404673, CHE-0911389 and 
CHE-0908218, and GF the Spanish MICINN grant FIS2007-61433
(co-financed FEDER) and the EU FP7 grant NMP4-SL-2011-266737.

\eject 

\end{document}